\documentclass [aps,12pt,a4paper]{revtex4}
\usepackage{graphicx}  
\usepackage{subfigure}
\usepackage{float}
\usepackage{dcolumn}   
\usepackage{bm}        
\usepackage{amssymb}   
\usepackage{natbib}

\begin{document}


\title{Charge regulation and local dielectric function in planar polyelectrolyte brushes}

\author{{\bf Rajeev Kumar} \footnote[1]{To whom any correspondence should be addressed, Email : kumarr@ornl.gov}}

\affiliation{National Center for Computational Sciences, Oak Ridge National Laboratory, Oak Ridge, TN-37831}

\author{{\bf Bobby G. Sumpter}} 

\affiliation{Center for Nanophase Materials Sciences, Oak Ridge National Laboratory, Oak Ridge, TN-37831}

\author{{\bf S. Michael Kilbey II}}
\affiliation{Center for Nanophase Materials Sciences, Oak Ridge National Laboratory, Oak Ridge, TN-37831 \\
\&\\
 Department of Chemistry, University of Tennessee, Knoxville, TN-37996}

\date{\today}
\begin{abstract}

\noindent Understanding the effect of
inhomogeneity on the charge regulation and 
dielectric properties, and how it depends on the
conformational characteristics of the macromolecules is a long-standing problem. 
In order to address this problem, we have developed a field-theory to study charge regulation 
and \textit{local} dielectric function in planar polyelectrolyte brushes. 
The theory is used to study a polyacid brush, which is comprised of 
chains end-grafted at the solid-fluid interface, in equilibrium with 
a bulk solution containing monovalent salt ions, 
solvent molecules and pH controlling acid. 
In particular, we focus on the effects of the concentration of added salt and pH of the bulk 
in determining the local charge and dielectric function. 
Our theoretical investigations reveal that the dipole moment of the ion-pairs formed as a result of counterion
adsorption on the chain backbones play a key role in affecting the local dielectric function. 
For polyelectrolytes made of monomers having dipole moments lower than the 
solvent molecules, dielectric decrement is predicted inside the brush region. However, the formation of 
ion-pairs (due to adsorption of counterions coming from the dissociation of added salt) 
more polar than the solvent molecules is shown to increase the magnitude of the dielectric function 
with respect to its bulk value. Furthermore, an increase in the bulk salt concentration 
is shown to increase the local charge inside the brush region. 
\end{abstract}

\maketitle

\section{Introduction}
Polyelectrolyte brushes\cite{alexander_brush,degennes_brush,milner_review,brush_review,ballauff_review,
ballauff_review_spherical} have been used extensively in the development of stimuli-sensitive materials due to their
ultra-sensitive response to different stimuli such as pH, temperature, solvent etc. A fundamental understanding of
the effect of each of these stimuli on the brush properties is highly desirable for a number of technologies.
Extensive research\cite{misra_varanasi_1,misra_varanasi_finite_extension,ross_pincus,zhulina_ionizable,birshtein_finite_extension,
biesheuvel_ionizable,szleifer_jps,genzer_macro,szleifer_macro_07,szleifer_prl,
orland_scft_brushes,kilbey07,kilbey08,witte_jpcb,
szleifer_pnas,szleifer_langmuir} aimed at elucidating the effect of different experimental variables on the response of 
polyelectrolyte brushes has been conducted.
However, in general, dielectric properties of polyelectrolytes\cite{mandel_jenard1,mandel_jenard2,
mandel_dielectric,colby_dielectric,minakata} elude a clear 
understanding and are
rarely studied using rigorous theoretical tools. 

Classic literature 
on the dielectric function\cite{debye_book,onsager_moments,dielectric,booth_1,booth_2,fulton_approaches} 
mainly deals with the small molecules randomly distributed in space, constituting a spatially homogeneous medium.
For a spatially uniform medium, in the absence of explicit charges, the dielectric function also turns out to be 
invariant in space and hence, the words ``dielectric constant" are used in place of ``dielectric function". 
For a single charge in a polar medium, it has been 
shown\cite{debye_book,onsager_moments,dielectric,booth_1,booth_2,fulton_approaches,
sandberg,freed_dielec1,freed_dielec2} 
that the dielectric function 
depends on the distance from the charge. The dielectric function is lowest next to the charge 
and monotonically approaches the dielectric 
constant of the medium. This effect is known as the dielectric saturation effect\cite{debye_book} 
resulting from the strong electric 
field next to the charge, which vanishes with an 
increase in the distance from the charge, reaching zero in the bulk.

The dielectric saturation effect has been shown\cite{debye_book} to be responsible for
the dielectric decrement when \textit{trace} amounts of
electrolytes like sodium chloride are added to polar solvents
such as water. In particular, a linear dependence of the dielectric decrement on the salt concentration
was proposed by Debye\cite{debye_book}. The linear
relation between the dielectric decrement and the salt concentration was also  observed in
experiments\cite{debye_book}.
However, the situation at large concentrations of salt is far more complex due to the interplay
between association/ion-pairing and short-range interactions.

Like electrolytes, polyelectrolytic systems are inherently inhomogeneous.
Understanding the effect of
inhomogeneity on the dielectric properties of the macromolecules is a long-standing problem.
To the best of our knowledge, there is no comprehensive theory for the \textit{local} 
dielectric function of polyelectrolytes, 
despite a large body of experimental work\cite{mandel_jenard1,mandel_jenard2,
mandel_dielectric,colby_dielectric}, which has shown that 
addition of polyelectrolytes to a solution containing polar solvent and salt 
leads to dielectric increment\cite{mandel_jenard1,mandel_jenard2,mandel_dielectric,colby_dielectric}. 
Minakata et al.\cite{minakata} attributed the \textit{overall} dielectric increment to the fluctuations of
the bound counterion charge at different sites. However, the theory by Minakata et al.\cite{minakata} 
ignores the conformational degrees of freedom and deals with \textit{rod-like} polyelectrolytes 
in an external electric field. 

From the above discussion, it is clear that the dielectric function 
depends on the number of charges and their distribution. 
This dependence of the dielectric function adds another 
complexity in the case of polyelectrolytes. The complexity arises because of 
the tendency of polyelectrolytes to regulate their charges\cite{manning,prabhu05,levin_review,
granick,kumar_condensation} on the basis of 
the local environment such as local pH, salt 
cocentration, etc.  In other words, one has to compute the local charge and the dielectric function 
in a self-consistent manner while minimizing the free energy. 
In this work, we have developed a quantitative description of the 
charge regulation and local dielectric function for polyelectrolytes. 
The theoretical formalism is quite general and we use it to 
study a planar polyelectrolyte brush in equilibrium with a 
solution bath containing monovalent salt ions, solvent molecules and an acid to control 
the pH of the bath. This particular set-up serves as a model system and is relevant for 
a number of experimental studies.
Typically, solvent molecules are treated as a
continumm without having any structure. Here, we treat each solvent molecule as having a permanent 
dipole moment of its own, which
is a reasonable model representation of a number of polar molecules. 
For a quantitative analysis of the dielectric function, which must depend
on the local
properties such as density, electrostatic potential and electric field, we have 
developed a field-theory taking into account the
dipolar interactions between charged and uncharged components.

This paper is organized as follows: the general formalism is presented in 
section ~\ref{sec:theory} and details about the application of the formalism to the 
polyelectrolyte brushes are presented in section ~\ref{sec:brushes}. 
Numerical results are presented in section ~\ref{sec:results}, and 
section ~\ref{sec:conclusions} contains our conclusions.

\section{Theory}\label{sec:theory}
\begin{figure}[h]  \centering
\vspace{0.2in}
\includegraphics[height=3.5in,width=3.5in]{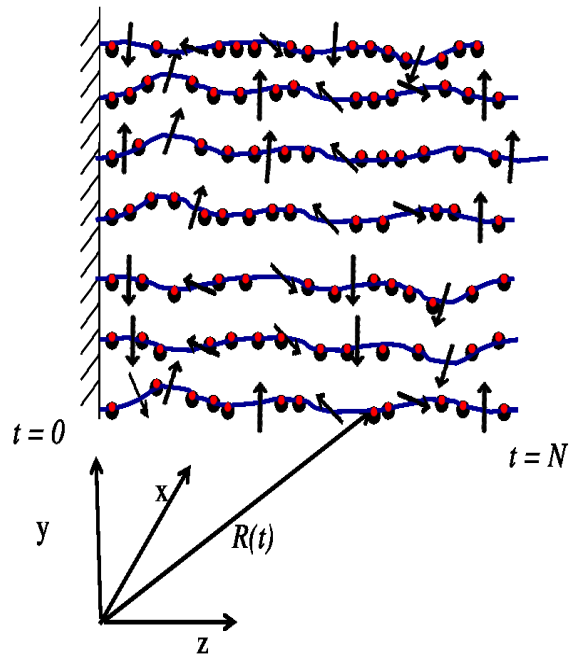}
\caption{Schematic of a planar polyelectrolyte brush containing 
monodisperse chains in equilibrium with an electrolyte 
solution (not shown explicitly here). The polyelectrolyte chains are modeled as 
continuous curves containing charges (represented by 
red dots) and electric dipoles (represented by arrows) resulting from counterion 
adsorption. The $\mbox{pH}$ regulating electrolyte solution is assumed to contain 
two kinds of counterions resulting into two kinds 
of dipoles on the chains.}
\label{fig:cartoon}
\end{figure}

\setcounter {equation} {0} 
We consider a planar polyelectrolyte brush formed by $n$ mono-disperse flexible polyacid chains 
(such as poly(methacrylic acid)), each having $N$ Kuhn
segments of length $l$. The chains are assumed to be uniformly grafted onto an uncharged substrate so that 
the grafting density (defined as the number of chains per square nanometer)
is $\sigma$ (see Fig. ~\ref{fig:cartoon}). For the field theoretical analysis\cite{fredbook,edwardsbook} described 
in this work, each polyelectrolyte chain
is represented by a continuous curve of length $Nl$, and an arc variable $t$
is used to represent any segment along the backbone so that $t \in [0,N]$. 
$t=0$ corresponds to the grafted end and $t=N$ represents the other end. 
To keep track of different grafted chains, subscript $\alpha$ is used so that $t_{\alpha}$
represents the contour variable along the backbone of $\alpha^{th}$ chain.
In the following, we use the notation
 $\mathbf{R}_{\alpha}(t_{\alpha})$ to represent the position vector for a 
particular segment,$t_{\alpha}$, along the $\alpha^{th}$ chain.

Counterion adsorption on the chains is studied using a two-state model. In this model, 
the entire population of counterions is divided based on whether they are ``free'' or ``adsorbed''. 
As the label implies, the ``free'' counterions are free to sample the whole space, while  the ``adsorbed'' 
counterions have less translational degrees of freedom and are bound to the chains.
In other words, each segment of a chain can be in either uncharged or charged state.
The adsorbed counterions/uncharged segments are treated as ``point'' dipoles. 
For each uncharged segment $t_\alpha$ along the $\alpha^{th}$ chain, an electric dipole of 
moment (in units of electronic charge, $e$) $\mathbf{p}_{\alpha}(t_\alpha)$ is assigned. 
Similarly, each solvent molecule is assigned an electric 
dipole moment and we use the notation $\mathbf{p}_k$ to represent 
the dipole moment of the $k^{th}$ solvent
molecule. Interactions between the charged and uncharged species are
taken into account by ion-dipole potentials. 

In this work, we focus on the effects of ``permanent dipole moments'' 
and don't allow variations in the magnitude of the electric dipole moments. 
However, we allow the dipoles to be aligned depending on the
local electric field. 
The physical origin of these electric dipoles is the difference in electronegativities of the atoms constituting
the dissociable group along the backbone. For example, acetic acid has a
dipole moment of $1.7$ Debye\cite{intermolecular_forces}.
It is well-known\cite{debye_book,onsager_moments,dielectric,booth_1} that finite 
polarizability of molecules leads to an enhancemnet of 
net electric dipole moments. We plan to include these effects in future work. 

We assume that the brush is in equilibrium with a bulk solution whose pH is controlled by 
the use of a buffer containing 
an acid and the precise concentrations of protons, $\mbox{H}^{+}$, cations from the added 
salt, $\mbox{B}^{+}$, and 
anions, $\mbox{A}^{-}$, which result from 
the dissociation of the acid, water molecules (protons and hydroxyl ions) 
and the salt, are known. For the sake of simplicity, we assume that the anions coming from 
the acid, water molecules and the salt are indistinguishable. 
Generalization of the theory presented here to take into account the 
specificity of anions is straightforward. 
However, we consider the specificity of cations (counterions for the brush) 
to study the effects of dipole moments of different kinds of ion-pairs and 
competitive counterion adsorption. 
Physically, $\mbox{H}^{+}$ originates from the dissociation of acidic sites on the polyelectrolytes, 
protonation of water molecules and the acid (for pH $< 7$). In this paper, we use the word 
``bulk'' quite frequently while referring to the region in space, where electrostatic potential and densities are 
spatially independent. One should not confuse the region just outside the brush region as the ``bulk'', because 
the electrostatic potential may be inhomogeneous even in monomer-free regions.  

The theory is developed in the Canonical ensemble 
 and the equilibrium with the bulk solution is considered by equating the 
chemical potentials of different components in the 
inhomogeneous brush and the homogeneous bulk solution. For the theoretical treatment carried out in 
the Canonical ensemble, we assume that there are $n_\gamma$ ions of kind $\gamma = H^+,B^+,A^-$. 
The number of counterions in the ``free'' and ``adsorbed'' states are represented by 
superscripts $f$ and $a$, respectively. 
Overall, the system is electroneutral and local electroneutrality is 
assumed for the bulk solution. To keep track of different kinds of ions, we define
$Z_\gamma$ as the valency (with sign) of the charged species $\gamma = H^+,B^+,A^-,p$, 
where the subscript $p$ represents the polymers.
For monovalent ions, $|Z_\gamma| = 1$ and for a polyacid, $Z_p = -1$. 

Following the mathematical procedure presented in Appendices A and B, we can write the 
partition function for the polyelectrolyte brushes as 
\begin{eqnarray}
       Z & = & \frac{1}{\Xi}\int D\left[\rho_p\right]\int  D\left[w_p\right]\int D\left[w_s\right]\int  D\left[\psi\right] 
\exp \left [-\frac{F_0}{k_BT} - \frac{H}{k_BT}\right ] \label{eq:parti_physical}
\end{eqnarray}
where
\begin{eqnarray}
      \frac{F_0}{k_B T} &=& \frac{F_a}{k_B T} + \frac{1}{2}\left[w_{pp} \rho_{po}nN + w_{ss}\rho_{so}n_s\right] 
- \ln \left[\Omega^{\sum_{\gamma'} n_{\gamma'}^f + n_{A^-} + n_s}\right], \quad \gamma' = H^+, B^+ 
\end{eqnarray}
and $F_a$ contains contributions coming from the adsorbed counterions. Explicitly, it is given by (see Appendix B)
\begin{eqnarray}
       \frac{F_a}{k_B T} & = & n_{B^+}^a\ln \mbox{K}_{B^+} - (nN-n_{H^+}^a)\ln \mbox{K}_{H^+}
- \ln \left[\frac{nN!}{n_{H^+}^a! n_{B^+}^a!(nN- n_{H^+}^a - n_{B^+}^a)!}\right] \nonumber \\
&& + \ln \left[n_{H^+}^f! n_{B^+}^f! n_s!\right] - (nN + n_s) \ln 4\pi \label{eq:fadsorbed}
\end{eqnarray}
In writing Eq. ~\ref{eq:fadsorbed}, we have defined an equilibrium 
constant, $\mbox{K}_{H^+}$, for the dissociation of acidic groups on the
polyelectrolyte chains by the thermodynamic
relation\cite{mcquarie} $\mbox{K}_{H^+} = \exp[-\left\{\mu_{COO^-}^o + \mu_{H^+}^o - \mu_{COOH}^o\right\}]$
where $\mu$ is the chemical potential (in units of $k_BT$, the Boltzmann constant times the absolute temperature) 
and superscript $o$ represents the limit of infinite dilution.
Similarly, we have defined another equilibrium constant $\mbox{K}_{B^+}$
for the binding of the counterions from
the salt by the relation
$\mbox{K}_{B^+} = \exp[-\left\{\mu_{COO^-}^o + \mu_{B^+}^o - \mu_{COO^-B^+}^o\right\}]$.

Also, 
\begin{eqnarray}
       \frac{H}{k_BT} &=& \chi_{ps}\int d\mathbf{r} \rho_p(\mathbf{r})\rho_{so}\left[1 - \frac{\rho_p(\mathbf{r})}{\rho_{po}}\right]  
 + i \int d\mathbf{r} \left[\psi_p(\mathbf{r}) - w_p(\mathbf{r})\right]\rho_p(\mathbf{r}) \nonumber \\
&& - \frac{1}{8\pi l_{Bo}}\int d\mathbf{r} \psi(\mathbf{r})\nabla_{\mathbf{r}}^2 \psi(\mathbf{r}) 
- i \int d\mathbf{r}\rho_{so}\left[1 - \frac{\rho_p(\mathbf{r})}{\rho_{po}}\right]\left[ w_s(\mathbf{r}) 
+ \ln \left[\frac{\sin\left(p_s |\nabla_\mathbf{r}\psi(\mathbf{r})|\right)}
{p_s |\nabla_\mathbf{r}\psi(\mathbf{r})|}\right]\right] \nonumber \\
&& - \sum_{\gamma' = H^+, B^+} n_{\gamma'}^f \ln Q_{\gamma'}\left\{\psi\right\} - n_{A^-} \ln Q_{A^-}\left\{\psi\right\} 
- n_s\ln Q_{s}\left\{w_s\right\} 
- \sum_{\alpha=1}^n Q_{p\alpha}\left\{w_p\right\} \label{eq:hami_physical} 
\end{eqnarray}
Here, $Q_\gamma,Q_s$ and $Q_{p\alpha}$ are the \textit{normalized} partition functions for a 
small ion of type $\gamma$, solvent molecule, and $\alpha^{th}$ grafted 
chain, respectively, given by
\begin{eqnarray}
      Q_{\gamma}\left\{\psi\right\} &=& \frac{1}{\Omega}\int d\mathbf{r} \exp\left[-i Z_\gamma \psi(\mathbf{r})\right] \quad \mbox{for}\, \gamma = H^+, B^+,A^-\\
      Q_{s}\left\{w_s\right\} &=& \frac{1}{\Omega}\int d\mathbf{r} \exp\left[-i w_s(\mathbf{r})\right] 
\end{eqnarray}
and
\begin{eqnarray}
      Q_{p\alpha}\left\{w_p\right\} &=& \int_{\mathbf{R}_\alpha(0)= \mathbf{r}_\alpha} D\left[\mathbf{R}_\alpha(t_\alpha)\right] \exp\left[-\frac{3}{2l^2}\int_{0}^N dt_\alpha 
\left(\frac{\partial \mathbf{R}_\alpha(t_\alpha)}{\partial t_\alpha}\right)^2 - i \int_{0}^{N} dt_\alpha w_p(\mathbf{R}_\alpha)\right] 
\end{eqnarray}
where $\mathbf{r}_\alpha$ is the position vector for the grafted end of the $\alpha^{th}$ chain. 

In these equations, $\psi(\mathbf{r})$ is the collective field introduced to decouple electrostatic interactions 
and it is the equivalent to the electrostatic 
potential. $w_{p}(\mathbf{r}), w_s(\mathbf{r})$ are the fields 
introduced to decouple short range interactions modeled by the Edwards' delta 
functional approach\cite{fredbook,edwardsbook}. $\rho_p(\mathbf{r})$ is the collective density 
variable. The Edwards' approach 
for using a delta function potential to model short range interactions along with the local 
incompressibility constraint naturally leads to appearance of the parameter $\chi_{ps}$ in the theory, defined in 
terms of excluded volume parameters $w_{pp},w_{ss}$ and $w_{ps}$ in Appendix B. 

Also, $l_{Bo} = e^2/\epsilon_o k_B T$ is the Bjerrum length in vacuum, 
where $e$ is the charge on an electron and $\epsilon_o$ being the 
permittivity of vacuum.
$\Omega$ is the total volume of the system and $p_s = |\mathbf{p}_k|$ is the 
magnitude of the dipole moment of each solvent molecule. $\rho_{po}$ and $\rho_{so}$ are the 
densities of pure polymers and solvent, respectively,  
and $\Xi$ is a normalizing factor defined in the Appendix B. 
The function $\psi_p$ appearing in Eq. 
~\ref{eq:hami_physical} is given by (see Appendix B)

\begin{eqnarray}
       \exp\left[- i \psi_p(\mathbf{r})\right] & = & (1 - \beta_{H^+} - \beta_{B^+}) \exp\left[-i Z_p \psi(\mathbf{r})\right] 
+ \sum_{\gamma' = H^+,B^+} \beta_{\gamma'} \left[\frac{\sin\left(p_{\gamma'} |\nabla_\mathbf{r}\psi(\mathbf{r})|\right)}
{p_{\gamma'} |\nabla_\mathbf{r}\psi(\mathbf{r})|} \right] \label{eq:psi_p}
\end{eqnarray}
where $\beta_{\gamma'}, \gamma' = H^+,B^+$ are the variational parameters characterizing 
the extent of counterion adsorption of type $\gamma'$ on the polyelectrolyte chains. 
These parameters need to be determined by the minimization of the free energy of the system. 
Furthermore, $p_{\gamma'}$ is the magnitude of the dipole moment of an ion-pair formed by 
the adsorption of counterion of type $\gamma'$. 

In the next section, we present the implications of the dipolar interactions on the 
properties of planar polyelectrolyte 
brushes in equilibrium with a solution containing monovalent salt.

\section{Planar polyelectrolyte brush in equilibrium with a solution bath} \label{sec:brushes}
\subsection{Saddle-point approximation}
An inhomogenous polyelectrolyte brush in equilibrium with a salty solution can be studied using the saddle-point 
approximation\cite{fredbook}. The approximation evaluates the functional integrals over the fields 
by the value of the integrand at the saddle-point. 
Optimizing the Hamiltonian given by Eq. ~\ref{eq:hami_physical} with respect to $w_p,w_s,\rho_p$ and $\psi$, 
respectively, we obtain 
\begin{eqnarray}
\rho_p^\star(\mathbf{r})  &=& -\sum_{\alpha=1}^n\frac{\delta \ln Q_{p\alpha}\left\{w_p^\star\right\}}{\delta w_p^\star(\mathbf{r})}\label{eq:monomer_den}\\
\rho_{so}\left[1-\frac{\rho_p^\star(\mathbf{r})}{\rho_{po}}\right]  &=& \frac{n_s \exp\left[-w_s^\star(\mathbf{r})\right]}
{\int d\mathbf{r}\exp\left[-w_s^\star(\mathbf{r})\right]}
\end{eqnarray}
\begin{eqnarray}
w_p^\star(\mathbf{r}) &=& \psi_p^\star(\mathbf{r}) 
+ \frac{\rho_{so}}{\rho_{po}} \left \{ w_s^\star(\mathbf{r}) + 
\ln \left[\frac{\sinh\left(p_s |\nabla_\mathbf{r}\psi^\star(\mathbf{r})|\right)}{p_s |\nabla_\mathbf{r}\psi^\star(\mathbf{r})|}\right] \right \}
+ \chi_{ps}\rho_{so}\left[1 - \frac{2\rho_p^\star(\mathbf{r})}{\rho_{po}}\right]
\end{eqnarray}

\begin{eqnarray}
\nabla_{\mathbf{r}}.\left[\frac{1}{l_B\left\{\rho_p^\star,\psi^\star\right \}} \nabla_{\mathbf{r}} \psi^\star(\mathbf{r})\right] &=& - 4\pi \left[\sum_{\gamma'=H^+,B^+} Z_{\gamma'} 
\rho_{\gamma'}^\star(\mathbf{r}) + Z_{A^-} \rho_{A^-}^\star(\mathbf{r}) + Z_p \beta^\star(\mathbf{r}) \rho_p^\star(\mathbf{r})\right] \label{eq:saddle_psi}
\end{eqnarray}
Here, we have used the notation $i w_p(\mathbf{r}) = w_p^\star(\mathbf{r}), i w_s(\mathbf{r}) = w_s^\star(\mathbf{r}),
i \psi(\mathbf{r}) = \psi^\star(\mathbf{r}), i \psi_p(\mathbf{r}) = \psi_p^\star(\mathbf{r})$ 
because the collective fields are purely imaginary\cite{fredbook} at the saddle point. 
Also, the collective density variables such as $\rho_p(\mathbf{r})$ at the saddle point are 
as $\rho_p^\star(\mathbf{r})$. $\rho_\gamma^\star(\mathbf{r})$ in Eq. ~\ref{eq:saddle_psi} 
is the number density of small ions of type $\gamma = H^+,B^+,A^-$, given by
\begin{eqnarray}
\rho_\gamma^\star(\mathbf{r}) &=& \frac{n_\gamma \exp\left[-Z_\gamma \psi^\star(\mathbf{r})\right]}{\int d\mathbf{r}\exp\left[-Z_\gamma \psi^\star(\mathbf{r})\right]} \label{eq:den_small_ions}
\end{eqnarray}
where $n_{\gamma} = n_{\gamma'}^f$ for $\gamma = H^+,B^+$. In other words, only the ``free'' counterions 
appear in Eq. ~\ref{eq:saddle_psi}. Furthermore, the effective Bjerrum length in Eq. ~\ref{eq:saddle_psi} is given by
\begin{eqnarray}
\frac{1}{l_B\left\{\rho_p^\star,\psi^\star\right \}}  &=& \frac{1}{l_{Bo}} 
+ 4\pi \left\{ \sum_{\gamma' = H^+,B^+}p_{\gamma'}^2 p_{\gamma'}^\star(\mathbf{r}) \rho_p^\star(\mathbf{r}) 
\bar{L}\left[p_{\gamma'} |\nabla_\mathbf{r} \psi^\star(\mathbf{r})|\right] \right .
\nonumber \\
&& \left . + p_s^2 \rho_{so}\left[1-\frac{\rho_p^\star(\mathbf{r})}{\rho_{po}}\right]\bar{L}\left[p_s |\nabla_\mathbf{r} \psi^\star(\mathbf{r})|\right]\right \}  \label{eq:saddle_lb}
\end{eqnarray}
where $\bar{L}(x) = L(x)/x$ so that $L(x) = \coth x - 1/x$ is the Langevin function.
Furthermore,
\begin{eqnarray}
\beta^\star(\mathbf{r}) &=& \frac{(1 - \beta_{H^+} - \beta_{B^+}) \exp\left[-Z_p\psi^\star(\mathbf{r})\right]}{\exp\left[-\psi_p^\star(\mathbf{r})\right]} \label{eq:fstar}\\
p_{\gamma'}^\star(\mathbf{r}) &=& \frac{\beta_{\gamma'}}{\exp\left[-\psi_p^\star(\mathbf{r})\right]}\frac{\sinh(p_{\gamma'} |\nabla_\mathbf{r} \psi^\star(\mathbf{r})|)}{p_{\gamma'} |\nabla_\mathbf{r} \psi^\star(\mathbf{r})|} \label{eq:pstar}
\end{eqnarray}
so that $\psi_p^\star$ is given by Eq. ~\ref{eq:psi_p} at the saddle point with the notation 
$i\psi_p(\mathbf{r}) = \psi_p^\star(\mathbf{r})$. Explicitly, it is given by 
\begin{eqnarray}
       \exp\left[- \psi_p^\star(\mathbf{r})\right] & = & (1 - \beta_{H^+} - \beta_{B^+}) 
\exp\left[-Z_p \psi^\star(\mathbf{r})\right] 
+ \sum_{\gamma' = H^+,B^+}\beta_{\gamma'} \left[\frac{\sinh\left(p_{\gamma'} |\nabla_\mathbf{r}\psi^\star(\mathbf{r})|\right)}{p_{\gamma'} |\nabla_\mathbf{r}\psi^\star(\mathbf{r})|} \right]
\end{eqnarray}
Note that $\beta^\star(\mathbf{r}) + \sum_{\gamma' = H^+,B^+} p_{\gamma'}^\star(\mathbf{r})  = 1$. 
From Eqs. ~\ref{eq:fstar} and ~\ref{eq:pstar} we can 
interpret that $\beta^\star(\mathbf{r})$ and $p_\gamma^\star(\mathbf{r})$ denote the probability of finding 
a charge and a dipole resulting from adsorption of counterion of type $\gamma$, respectively, 
at location $\mathbf{r}$.

Invoking the relation between the Bjerrum length and local dielectric function, 
$l_B\left\{\rho_p^\star,\psi^\star\right \} = e^2/\epsilon\left\{\rho_p^\star,\psi^\star\right \}k_B T$ ,
we obtain
\begin{eqnarray}
\frac{\epsilon\left\{\rho_p^\star,\psi^\star\right \}}{\epsilon_{o}}  &=& 1+ 
4\pi l_{Bo}\left[\sum_{\gamma' = H^+,B^+}p_{\gamma'}^2 p_{\gamma'}^\star(\mathbf{r})
\rho_p^\star(\mathbf{r})\bar{L}\left[p_{\gamma'} |\nabla_\mathbf{r} \psi^\star(\mathbf{r})|\right]
\right . \nonumber \\
&& \left . + p_s^2 \rho_{so}\left[1-\frac{\rho_p^\star(\mathbf{r})}{\rho_{po}}\right]\bar{L}\left[p_s |\nabla_\mathbf{r} \psi^\star(\mathbf{r})|\right]\right]\label{eq:dielectric_saddle}
\end{eqnarray}
It is worth noting that the functional form of the field
dependent dielectric function in Eq. ~\ref{eq:dielectric_saddle} is qualitatively similar to the expression
derived in the classic papers by Onsager\cite{onsager_moments} and Booth\cite{booth_1,booth_2}.
Noting that $\bar{L}(x)$ decreases with an increase in $x$ and $\sinh x/x$ (appearing in $p_{\gamma'}^\star$, 
cf. Eq. ~\ref{eq:pstar}) increases with an increase in $x$, 
the local field can lead to dielectric decrement or increment depending on the relative 
strength of these counteracting effects. In other words, the first effect (also arising 
in the Langevin-Debye model\cite{debye_book} and known as 
the dielectric saturation effect) tends to lower 
the dielectric function. On the other hand, the second effect resulting from the counterion adsorption 
tend to increase it. 

An important insight into the effect of electric dipoles on the thermodynamics
of the polyelectrolyte brushes is obtained if one considers the weak coupling limit 
for the electric dipoles (i.e., the limit of weak dipoles or weak local electric fields) so that
$p_{\gamma'} |\nabla_\mathbf{r}\psi^\star(\mathbf{r})|\rightarrow 0$, $p_s|\nabla_\mathbf{r}\psi^\star(\mathbf{r})| 
\rightarrow 0$.
In these limits we can either use the approximation $\ln \left[\sin x /x\right] \simeq -x^2/6$ in the expression
for the partition function or $L(x) \rightarrow x/3$ for $x\rightarrow 0$ in Eq. ~\ref{eq:saddle_lb}.
Using either of the approximations, it can be shown that the electric dipoles renormalize the Bjerrum length 
of the medium by the relation
\begin{eqnarray}
\frac{1}{l_B\left\{\rho_p^\star,\psi^\star\right \}} \equiv \frac{1}{l_B\left\{\rho_p^\star\right \}} 
&=& \frac{1}{l_{Bo}} + \frac{4\pi}{3}\left[\sum_{\gamma' = H^+,B^+}\left(p_{\gamma'}^2 
\beta_{\gamma'}\right) \rho_p^\star(\mathbf{r}) 
+ p_s^2 \rho_{so}\left[1 - \frac{\rho_p^\star(\mathbf{r})}{\rho_{po}}\right]\right] 
\label{eq:bjerrum_weak}
\end{eqnarray}
This allows us to write $\epsilon(\mathbf{r}) = \epsilon_{p}\phi_p(\mathbf{r}) + \epsilon_s \phi_s(\mathbf{r})$,
where $\phi_p(\mathbf{r}) = \rho_p^\star(\mathbf{r})/\rho_{po}$ 
and $\phi_s(\mathbf{r}) = 1 - \rho_p^\star(\mathbf{r})/\rho_{po}$ are
the volume fractions of the monomers and solvent molecules, respectively. Also,
$\epsilon_p$ and $\epsilon_s$ are the dielectric constants of the pure components, given by
$\epsilon_p/\epsilon_{o} - 1 = 4\pi l_{Bo}\sum_{\gamma' = H^+,B^+}
\left(\beta_{\gamma'}p_{\gamma'}^2\right)\rho_{po}/3$ and by 
$\epsilon_s/\epsilon_{o} - 1 = 4\pi l_{Bo}p_s^2\rho_{so}/3$.
In other words, the linear mixing rule based
on the volume fractions can be used to estimate the local dielectric function
of the medium in the weak-coupling limit.

Furthermore, by optimizing the Hamiltonian given by Eq. ~\ref{eq:hami_physical} 
with respect to the variational parameter 
$\beta_{H^+}$, we obtain
\begin{eqnarray}
\log_{10} \left[\frac{\beta_{H^+}}{1- \sum_{\gamma'}\beta_{\gamma'}} \frac{\int d \mathbf{r}\exp\left[-Z_{H^+}
\psi^\star(\mathbf{r})\right]}{n_{H^+}^f}\right]
- \mbox{pK}_{H^+} && \nonumber \\  
+  \frac{1}{2.303 nN}\int d\mathbf{r} \rho_p^\star(\mathbf{r})
\left[\frac{\beta^\star(\mathbf{r})}{1- \sum_{\gamma'}\beta_{\gamma'}} - \frac{p_{H^+}^\star(\mathbf{r})}
{\beta_{H^+}}\right] &=& 0 \label{eq:charging_para}
\end{eqnarray}
where $\gamma' = H^+, B^+$. In deriving Eq. ~\ref{eq:charging_para}, 
we have used the fact that $n_{H^+}^f = n_{H^+}^{total} - \beta_{H^+} nN$. 
Also, we have defined $\mbox{pK}_{H^+} = -\log_{10}\mbox{K}_{H^+}$.
The last term in Eq. ~\ref{eq:charging_para} captures the 
effects of finite monomer density and inhomogenity on the 
charge regulation in the brushes. This term can be interpretted as 
the shift in $\mbox{pK}_{H^+}$ and depends on the local environment. 
However, sign of the shift depends on the relative contributions of the charges and dipole. 

A similar equation is obtained for $\beta_{B^+}$, which can be written by 
replacing $H^+$ by $B^+$ in Eq. ~\ref{eq:charging_para}. For the binding of the counterions from 
the salt, we define $\mbox{pK}_{B^+} = -\log_{10}\mbox{K}_{B^+}$.
Eq. ~\ref{eq:charging_para} equates the chemical potentials of the counterions in the ``free'' state and the 
``adsorbed'' state. In the next section, we present the details about the treatment of grafted ends in the theory.

\subsection{Treatment of grafted ends: uniformly grafted brush}
The functional derivative of the chain partition function with an end grafted 
at $\mathbf{R}_\alpha(0) = \mathbf{r}_\alpha$ on the substrate 
allows us to write Eq. ~\ref{eq:monomer_den} as 
\begin{eqnarray}
\rho_p^\star(\mathbf{r}) &=& \sum_{\alpha=1}^n \frac{\int_0^N dt q_{\mathbf{r}_\alpha}(\mathbf{r},t)\bar{q}(\mathbf{r},N-t)}
{\int d\mathbf{r} q_{\mathbf{r}_\alpha}(\mathbf{r},t)\bar{q}(\mathbf{r},N-t)}, \label{eq:den_q}
\end{eqnarray}
where $\bar{q}(\mathbf{r},N-t)$ satisfies 
\begin{eqnarray}
\frac{\partial \bar{q}(\mathbf{r},t')}{\partial t'} &=& \left[\frac{l^2}{6} \nabla_\mathbf{r}^2 
- w_p^\star(\mathbf{r})\right]\bar{q}(\mathbf{r},t') \label{eq:mod_diff}
\end{eqnarray}
with the condition $\bar{q}(\mathbf{r},0) = 1$ for $t'= N-t = 0$. Similarly, 
$q_{\mathbf{r}_\alpha}(\mathbf{r},t)$ satisfies the same equation but with 
the initial condition $q_{\mathbf{r}_\alpha}(\mathbf{r},0) = \delta(\mathbf{r} - \mathbf{r}_\alpha)$. 

Noting that the denominator in Eq. ~\ref{eq:den_q} is independent of $t$, 
Muller\cite{muller_algo} has developed an efficient algorithm to 
solve these equations 
in three dimensional space. Following Muller, we choose $t=0$ for the denominator in Eq. ~\ref{eq:den_q} 
and use the initial conditions for $q$ and $\bar{q}$ to write 
Eq. ~\ref{eq:den_q} as 
\begin{eqnarray}
\rho_p^\star(\mathbf{r}) &=& \int_0^N dt \hat{q}(\mathbf{r},t)\bar{q}(\mathbf{r},N-t) \label{eq:den_muller}
\end{eqnarray}
where 
\begin{eqnarray}
\hat{q}(\mathbf{r},t) &=& \sum_{\alpha=1}^n \frac{q_{\mathbf{r}_\alpha}(\mathbf{r},t)}
{\bar{q}(\mathbf{r}_\alpha,N)}  \simeq \sigma \int d\mathbf{r}_{2d} \frac{q_{\mathbf{r}_\alpha}(\mathbf{r},t)}
{\bar{q}(\mathbf{r}_\alpha,N)}\label{eq:qhat_muller}
\end{eqnarray}
Here, $\int d\mathbf{r}_{2d}$ is the integral over the grafting points. 
The later transformation to go from a discrete sum over all the chains to the quadrature over the grafting points is 
strictly valid in the thermodynamic limit so that 
$n\rightarrow \infty, A ($=\mbox{area of the grafting plane}$) \rightarrow \infty$ and $\sigma = n/A$ is finite. 
Due to the linear relation between $\hat{q}$ and $q_{\mathbf{r}_\alpha}$, $\hat{q}$ also satisfies Eq. ~\ref{eq:mod_diff}, 
but with the initial condition
\begin{eqnarray}
\hat{q}(\mathbf{r},0) &=& \frac{\sigma \delta(z-\delta)}
{\bar{q}(\{x,y,\delta\},N)} \label{eq:qhat_muller_init}
\end{eqnarray}
where we have written $\mathbf{r}_\alpha = \{x_\alpha,y_\alpha,\delta\}$ in Cartesian co-ordinates so that $\delta \rightarrow 0$ is the 
parameter used to put the anchoring point at an infinitesimal distance above the substrate. 

The above transformation also allows us to write the sum over the chain partition functions in the Hamiltonian (cf. Eq. ~\ref{eq:hami_physical}) 
in a manageable form, given by 
\begin{eqnarray}
\sum_{\alpha=1}^n \ln Q_{p\alpha}\left\{w_p\right\} &\simeq& \sigma \int d\mathbf{r}_{2d}\ln \left[\bar{q}(\{\mathbf{r}_{2d},\delta\},N)\right] \\
   &=& \sigma A\ln \left[\bar{q}(\delta,N)\right]
\end{eqnarray} 
The last expression is valid in the special case when $\bar{q}(\mathbf{r},N)$ is a function of 
the distance from the substrate i.e., in the absence 
of lateral inhomogeneities. We call this particular case the one-dimensional brush because 
the densities of different components are 
dependent only on the distance from the substrate. This is the case for high grafting densities in good solvents. 

\subsection{Free energy with respect to the salty solution}
For a polyelectrolyte brush in equilibrium with a solution containing monovalent salt, we have to equate the 
chemical potentials 
of all the components that can be exchanged between the bulk solution and the brush region. Also, at equilibrium, 
the osmotic pressure must be the same everywhere. However, due to the incompressibility 
constraint, equating the chemical potentials of the salt ions and solvent molecules is sufficient to define the 
equilibrium state of the system. Furthermore, it is assumed that the densities of different components are known in the bulk solution 
and we use them as parameters in the study here. 

Using the thermodynamic relation between the free energy and chemical potentials in the canonical ensemble 
(i.e., $\mu_j = (\partial F/\partial n_j)_{\Omega,T} $), we get 
\begin{eqnarray}
\mu_\gamma &=& \ln \left[\frac{n_\gamma}{\int d\mathbf{r} \exp\left[-Z_\gamma \psi^\star(\mathbf{r})\right]}\right] 
= \ln \rho_{\gamma}(\infty) + Z_\gamma \psi^\star(\infty) \label{eq:chem_ions}
\end{eqnarray}
for the ``free'' ions, $\gamma  = H^+,B^+,A^-$ so that $n_{\gamma} = n_{\gamma}^f$ 
for $\gamma  = H^+,B^+$. In writing the second equation on the right in Eq. ~\ref{eq:chem_ions}, we 
have used Eq. ~\ref{eq:den_small_ions} 
and the boundary condition that the density of the ``free'' ions is given by $\rho_{\gamma} (\infty)$ 
in the bulk (designated by $\mathbf{r} \rightarrow \infty$). Also, $\psi^\star(\infty)$ is the 
value of $\psi^\star(\mathbf{r})$ in the bulk. The values of $\beta_{\gamma'}$ are 
determined by using Eq. ~\ref{eq:chem_ions} 
along with Eq. ~\ref{eq:charging_para}. An important insight into the effect of 
salt concentration on the charge regulation can be obtained by taking the limit of 
$\rho_p^\star(\mathbf{r}) = 0$ (i.e., the limit of vanishing
grafting density) in Eq. ~\ref{eq:charging_para}. 
Taking the limit of zero monomer 
density in Eq. ~\ref{eq:charging_para} and using Eq. ~\ref{eq:chem_ions}, we can write 
\begin{eqnarray}
\beta_{H^+} &=& \frac{1}{1 + 10^{\mbox{pH}-\mbox{pK}_{H^+}}\left[e^{-Z_+ \psi^\star(\infty)/2.303} 
+ \rho_{B^+}(\infty)10^{\mbox{pK}_{B^+}}\right]} \label{eq:binding_bulk1}\\
\frac{\beta_{B^+}}{\beta_{H^+}} &=& \rho_{B^+}(\infty)10^{\mbox{pH} - \mbox{pK}_{H^+} + \mbox{pK}_{B^+}} \label{eq:binding_bulk2}
\end{eqnarray}
where we have defined pH of the bulk as $\mbox{pH} = -\log_{10}\left[\rho_{H^+}(\infty)\right]$ and 
$Z_+ = |Z_{H^+}| = |Z_{B^+}|$ is the valency of counterions ($=1$ for the monovalent
ions considered in this work). $\rho_{B^+}(\infty)$ is the concentration of the cations 
coming from the added salt in the bulk.

Eqs. ~\ref{eq:binding_bulk1} and ~\ref{eq:binding_bulk2} 
capture the physics of competitive counterion adsorption on the polyelecrolyte chains
in the limit of vanishing monomer density. Eqs. ~\ref{eq:fstar} and 
~\ref{eq:pstar} capture the same in an inhomogeneous medium at finite monomer density.  
In the next section, we'll demonstrate that $\beta^\star(\mathbf{r})$, which determines the charge regulation in 
polyelectrolyte brushes, is highly sensitive to the competitive 
counterion adsorption captured by Eqs. ~\ref{eq:binding_bulk1} and ~\ref{eq:binding_bulk2}.
In the particular case, when the counterions from the added salt don't adsorb on the 
polyelectrolyte chains, $\mbox{pK}_{B^+} \rightarrow -\infty$ and $\beta_{B^+} \rightarrow 0$ 
and $\beta_{H^+} = 1/\left[1 + 10^{\mbox{pH}-\mbox{pK}_{H^+}}e^{-Z_+ \psi^\star(\infty)/2.303}\right]$. 
For this case, the degree of adsorption of the counterions from the 
salt ($\beta_{B^+} = 0$) and protons ($\beta_{H^+}$) become 
independent of the bulk salt concentration. However, in general, $\beta_{H^+}$ 
decreases and $\beta_{B^+}$ increases with an increase in the bulk salt concentration.  
Furthermore, in general, the dipoles formed as a result of the adsorption of counterions from the salt
have higher moments in comparison to those formed by the adsorption of protons.
In the next section, this difference in the dipole moments is shown to play
an important role in affecting the dielectric function within a polyelectrolyte brush.

Similarly, the chemical potential of solvent molecules is given by 
\begin{eqnarray}
\mu_s &=& \frac{w_{ss}\rho_{so}}{2} - \ln 4\pi + \ln \left[\frac{n_s}{\int d\mathbf{r} \exp\left[-w_s^\star(\mathbf{r})
\right]}\right] = \frac{w_{ss}\rho_{so}}{2} - \ln 4\pi + \ln \rho_{so} + w_s^\star(\infty).
\end{eqnarray}

Taking the homogeneous bulk solution as the reference frame, we can write the free energy at 
the saddle-point as $F^\star = F_{ref}^\star + \Delta F^\star$,
where $F_{ref}^\star = \sum_{\gamma = H^+,B^+,A^-,s} \mu_\gamma n_\gamma$. 
In this work, we are interested in the effect of inhomogeneities on the properties of 
polyelectrolyte brushes and $\Delta F^\star$ is the contribution to the free energy 
arising from the inhomogeneous distributons of fields. Because the free energy is defined with in an 
arbitrary constant, we subtract a constant to ensure that 
$\Delta F^\star = 0$ in the absence of inhomogeneities.  
Explicitly, $\Delta F^\star$ is given by 
\begin{eqnarray}
\Delta F^\star &=& F_{self} + F_{chemical} + F_w + F_{ions} + F_{solvent} + F_e + F_{poly},
\end{eqnarray}
where $F_{self}/k_B T = [w_{pp} \rho_{po}/2 - \ln 4\pi ]nN$ and 
\begin{eqnarray}
F_{chemical} &=& nN\left[\beta_{B^+}\ln K_{B^+} - (1-\beta_{H^+})\ln K_{H^+}\right] \nonumber \\
&& + nN\left[\sum_{\gamma'}\beta_{\gamma'}
\ln \beta_{\gamma'} + \left(1-\sum_{\gamma'}\beta_{\gamma'}\right)\ln\left[1-\sum_{\gamma'}\beta_{\gamma'}\right]\right], 
\end{eqnarray}
where $\gamma' = H^+,B^+$ and
\begin{eqnarray}
F_w &=& \chi_{ps}\int d\mathbf{r}\rho_p^\star(\mathbf{r})\rho_s^\star(\mathbf{r}) \\
F_{ions} &=& \sum_{\gamma = H^+,B^+,A^-} \int d\mathbf{r} \left[\rho_\gamma^\star(\mathbf{r})
\ln \left[\frac{\rho_\gamma^\star(\mathbf{r})}{\rho_{\gamma o}}\right]
 - \rho_\gamma^\star(\mathbf{r}) + \rho_{\gamma o}\right ] \\
F_{solvent} &=& \int d\mathbf{r} \left[\rho_s^\star(\mathbf{r})\ln \left[\frac{\rho_s^\star(\mathbf{r})}{\rho_{s o}}
\right] - \rho_s^\star(\mathbf{r}) + \rho_{s o}\right ]\\
F_e &=&  \int d\mathbf{r} \left[\sum_{\gamma = H^+,B^+,A^-}Z_{\gamma} 
\rho_{\gamma}^\star(\mathbf{r})\psi^\star(\mathbf{r}) 
+ \psi_p^\star(\mathbf{r})\rho_p^\star(\mathbf{r})\right ]
+ \frac{1}{8\pi l_{Bo}}\int d\mathbf{r} \psi^\star(\mathbf{r})\nabla_{\mathbf{r}}^2 \psi^\star(\mathbf{r}) \nonumber\\
&& - \int d\mathbf{r} \rho_s^\star(\mathbf{r}) \ln \left[\frac{\sinh \left(p_s|\nabla_\mathbf{r}\psi^\star(\mathbf{r})|\right)}{\left(p_s|\nabla_\mathbf{r}\psi^\star(\mathbf{r})|\right)}\right]\\
F_{poly} &=& -\sigma \int d\mathbf{r}_{2d}\ln \left[\bar{q}(\{\mathbf{r}_{2d},\delta\},N)\right] - i\int d\mathbf{r} \rho_p^\star(\mathbf{r})w_p^\star(\mathbf{r})
\end{eqnarray}
where $\rho_s^\star(\mathbf{r}) = \rho_{so}\left[1 - \rho_p^\star(\mathbf{r})/\rho_{po}\right]$.

\subsection{Numerical methods}
We have solved the non-linear set of equations assuming lateral homogeneity in the planar polyelectrolyte brush. 
For the brush in equilibrium with the solution bath, the above equations are written in 
dimensionless form and 
the direction prependicular to the grafting plane is assigned as the $z$ axis in Cartesian coordinates. 
All of the quantities having dimensions of 
length are made dimensionless by dividing them by $R_{go} = (Nl^2/6)^{1/2}$. 
Numerical calculations ensuring that the electric field far from the 
brush region approaches zero, demand a box length as high as $100 R_{go}$, 
especially in low salt conditions such as $1$ millimolar (mM). For the highest bulk salt concentration 
of $100$ mM, a box length of $20 R_{go}$ is found to be sufficient. 
However, the monomer density decays to zero within a distance of $10-15 R_{go}$ from the grafting 
surface for all the cases investigated in this work. Numerical results presented in this work have been 
obtained by taking the maximum box length of $100 R_{go}$ with $512$ grid points. 
  
The modified diffusion equation represented by Eq. ~\ref{eq:mod_diff} and the Poisson-Boltzmann equation 
(Eq. ~\ref{eq:saddle_psi}) have been solved by using implicit-explicit scheme known as 
the extrapolated gear method\cite{numerics_imex,numerics_badalassi}. 
In order to use the extrapolated gear method, we rewrite 
Eq. ~\ref{eq:saddle_psi} in the form
\begin{eqnarray}
\frac{\partial \psi^\star(\mathbf{r})}{\partial \bar{t}} &=& \nabla_{\mathbf{r}}.\left[\frac{1}{l_B\left\{\rho_p^\star,\psi^\star\right \}} \nabla_{\mathbf{r}} \psi^\star(\mathbf{r})\right] +  4\pi \rho_e^\star(\mathbf{r}) \label{eq:saddle_psi_2}
\end{eqnarray}
where $\bar{t}$ is a \textit{fictitious} time and $\rho_e^\star(\mathbf{r}) = \sum_{\gamma=H^+,B^+,A^-} Z_\gamma 
\rho_\gamma^\star(\mathbf{r}) + Z_p \beta^\star(\mathbf{r}) \rho_p^\star(\mathbf{r})$ 
is the local charge density. The steady state solution of 
Eq. ~\ref{eq:saddle_psi_2} is Eq. ~\ref{eq:saddle_psi}. 
Before implementing the extrapolated gear method in one dimension, we rewrite Eq. ~\ref{eq:saddle_psi_2} as 
\begin{eqnarray}
\frac{\partial \psi^\star(z)}{\partial \bar{t}} &=& \frac{\lambda_{max}}{2}\frac{\partial^2 \psi^\star(z)}
{\partial z^2} + \frac{\partial }{\partial z}\left[\left[\frac{1}{l_B\left\{\rho_p^\star,\psi^\star\right \}} -\frac{\lambda_{max}}{2}\right]\frac{\partial \psi^\star(z)}{\partial z}\right] +  4\pi \rho_e^\star(z) \label{eq:saddle_psi_3}
\end{eqnarray}
where $\lambda_{max} = 1/[l_B\left\{\rho_p^\star,\psi^\star\right \}]_{min}$ is chosen as the maximum value 
of inverse of the local Bjerrum length at each time step. 
In the extrapolated gear method, the first term on the right hand side of Eq. ~\ref{eq:saddle_psi_3} 
is treated implicitly and the rest is treated explicitly. The particular 
choice of $\lambda_{max}$ is motivated by the studies on the phase field models\cite{numerics_badalassi}, 
which require the solution of similar equations. An explicit scheme is used for the initialization of the extrapolated 
gear method.

An equation similar to Eq. ~\ref{eq:saddle_psi_3} is written while solving 
Eq. ~\ref{eq:mod_diff} with the choice $\lambda_{max} = 1/2$.
Furthermore, Crank-Nicholson scheme is used to initialize 
the gear method for the modified diffusion equation. 
Dirichlet boundary conditions are used for $\hat{q}$ and $\bar{q}$ 
at the substrate i.e., $\hat{q}(z=0,t) = \bar{q}(z=0,t) = 0 $ for all values of $t$. 
The grafted ends are displaced to the first grid point in the solution so that 
$\delta = 0.19R_{go}$ for a box length of $100R_{go}$ with $512$ grid points.
Also, steps of $10^{-4}$ and $10^{-3}$ are used for the time stepping 
while solving the Poisson-Boltzmann and the modified diffusion equations, 
respectively. 

In the absence of external electric fields, we have chosen $\psi^\star(\infty) = 0$.
Starting from an initial guess for the fields ($w_p^\star,\psi^\star$) and charge parameters 
($\beta_{H^+},\beta_{B^+}$), 
we use the extrapolated gear method to 
compute $\psi^\star$ at the next time step, which is used as a guess in the iterative scheme. 
On the other hand, the guessed and the computed values for 
$w_p^\star,\beta_{H^+}$ and $\beta_{B^+}$ are mixed using the 
simple mixing scheme\cite{fredbook} to develop a new guess for the next iteration. 
Random numbers have been used as initial guesses for the fields 
$w_p^\star$ and $\psi^\star$. 
However, Eqs. ~\ref{eq:binding_bulk1} and ~\ref{eq:binding_bulk2} are 
used as initial guesses for $\beta_{H^+}$ and $\beta_{B^+}$, respectively. 
The iterative procedure is continued until the free energy 
of the brush (with respect to the solution bath) doesn't change within $10^{-8}$. 

\section{Results} \label{sec:results}
Although the theory is quite general, in the following we have applied it to study a brush made of chains 
bearing acidic groups having $\mbox{pK}_{H^+} = 4.66$, which mimics poly(methacrylic acid)\cite{handbook}. 
The motivation behind this choice is the availability of data for the stability constants\cite{stability_constants1} of 
monovalent salts for this system and on-going experiments. For monovalent salts such as 
sodium chloride (NaCl) and lithium chloride (LiCl), it is 
found\cite{stability_constants1,stability_constants2} that $\mbox{pK}_{Na^+} \simeq \mbox{pK}_{Li^+} = 0.28$. 
Furthermore, we present the results 
for $p_{H^+} = 0.035 \mbox{nm} \equiv 1.7$ Debye  and $p_s = 0.1003 \mbox{nm} \equiv 4.8$ Debye 
mimicking weakly acidic groups such as 
methacrylic acid\cite{intermolecular_forces} 
and water, respectively, at the room temperature. 
The temperature is fixed in the calculations by the choice of $l_{Bo} = 56 \mbox{nm}$, which corresponds to 
room temperature.  
The choice of $p_s = 0.1003$ nm is motivated by the fact that 
the dielectric constant of water is $80$ at room temperature. Note that the value of $p_s$ 
chosen for the numerical work here is little higher than the actual dipole moment of water ($=1.85$ Debye 
in the gas phase\cite{intermolecular_forces}). 
The origin of this lies in the neglect of hydrogen bonding and induced dipole moment effects in the theory. 
Overcoming this limitation of the theory is an interesting direction for future research. 
Also, we have taken $l = 0.3103$ nm, which corresponds to $\rho_{so} = \rho_{po} = 1/l^3 \equiv 1 \mbox{g/cm}^3$ 
i.e., the density of water at room temperature. Note that 
the choice of $l = 0.3103$ nm is also close to the Kuhn segment length\cite{kuhn_pmaa} of 
poly(methacrylic acid). We have solved the equations at the saddle-point in three dimensional 
space for $N=100$ and our preliminary studies show that a grafting density of $\sigma R_{go}^2 = 1$ 
for $\chi_{ps}l^3 = 0.6$ is high enough to ensure that there are no lateral inhomogeneities. 
Note that the choice of $\chi_{ps}l^3 = 0.6$ represents a slightly poor-solvent 
quality for the polyelectrolyte chain backbone, which is definitely the case for 
poly(methacrylic acid) in an aqueous medium. All the results presented below are for $N=100$ and 
$\chi_{ps}l^3 = 0.6$.

Using these set of parameters relevant for poly(methacrylic acid) in an aqueous medium, 
we have varied the salt concentration and the pH of the bulk solution. To study the effect of 
adsorption of counterions coming from the salt, we have considered two cases. The first corresponds to 
$\mbox{pK}_{B^+} = 0.28$ and the second corresponds to $\mbox{pK}_{B^+} \rightarrow -\infty$, 
which represents the 
scenario in which the counterions from the salt do not adsorb on the chains. For the first case, where 
counterions from the salt can also adsorb on the acidic chains, we have taken $p_{B^+} = 0.125$ nm 
(corresponding to dipole moment of $6$ Debye for the sodium carboxylate (-COONa) group\cite{arai_eisenberg}).
 
\subsection{Effects of the bulk salt concentration}
\subsubsection{Monomer density profiles, charge regulation and the dielectric function}
\begin{figure}[ht]
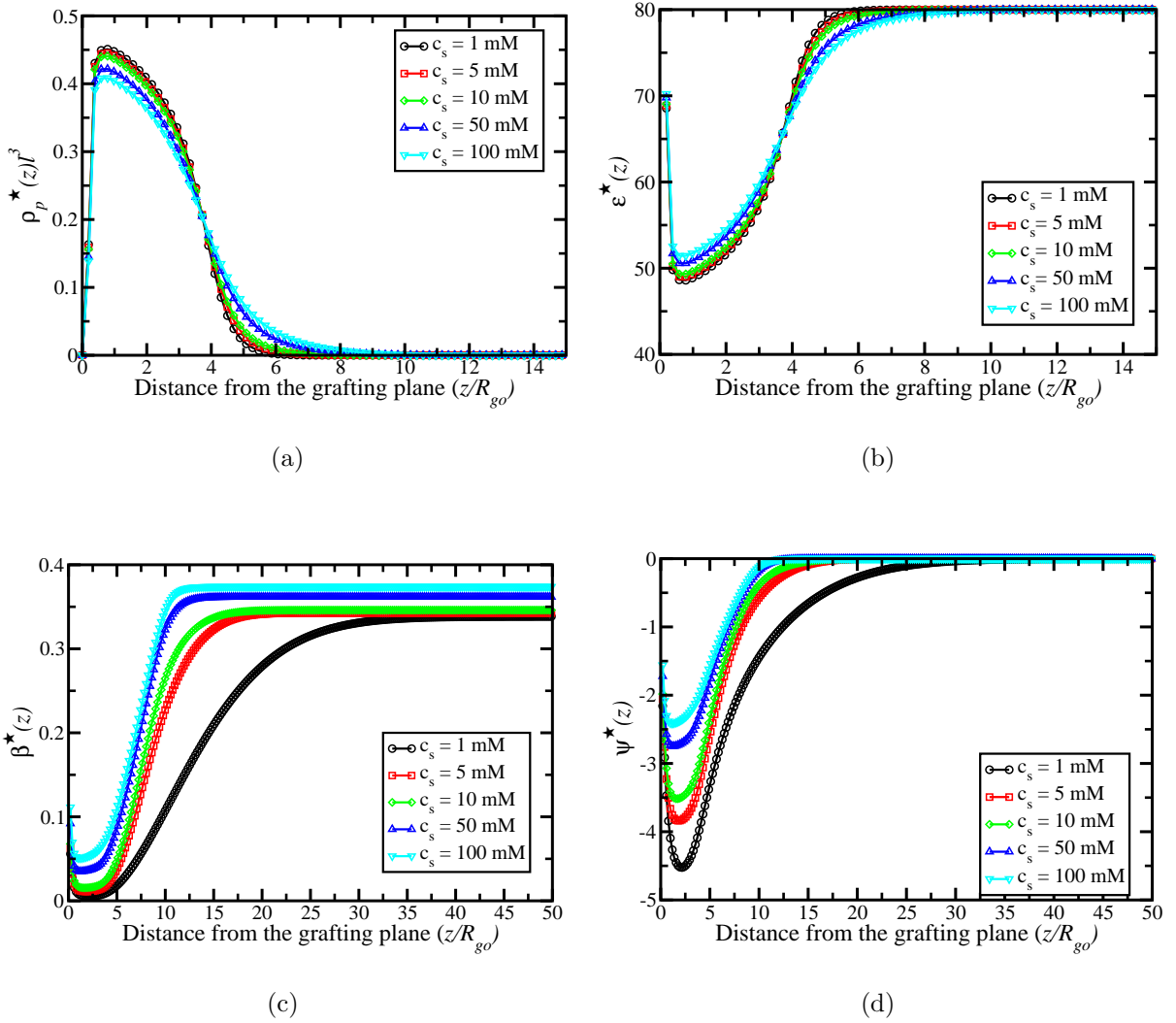
  \centering
\vspace{0.2in}
\subfigure[]{\includegraphics[width=3in]{cs_effect_monomer_wsolv1.eps}}
\hspace{0.1in}\subfigure[]{\includegraphics[width=3in]{cs_effect_dielectric_wsolv1.eps}}\vspace{0.3in}\\
\subfigure[]{\includegraphics[width=3in]{cs_effect_ionization_wsolv1.eps}}
\hspace{0.1in}\subfigure[]{
\includegraphics[width=3in]{cs_effect_elec_potential_wsolv1.eps}}
\caption{Effect of the bulk salt concentrations on properties of the
polyacidic brush in the case where counterions from the salt don't adsorb on the chains.
These results are obtained for bulk pH $= 5$ and $\mbox{pK}_{B^+} \rightarrow -\infty$.
Other parameters are presented in the main text. Figs. (a),(b),(c) and (d) correspond to 
the monomer density ($\rho_p^\star(z)$), local dielectric 
function ($\epsilon^\star(z)$), probability of finding a monomer in a charged state ($\beta^\star(z)$) and 
the electrostatic potential ($\psi^\star(z)$), respectively. The saddle-point equations are solved 
for $z_{max}/R_{go} = 100$ and the data is cut at $z/R_{go} = 15$ and $50$ to highlight the important
features for the Figs. (a)-(b) and (c)-(d), respectively.}
\label{fig:cseffect_ph5_1dip}
\end{figure}

In Fig. ~\ref{fig:cseffect_ph5_1dip} we present the results obtained for the polyacidic 
brushes when the bulk salt concentration,$c_s = \rho_{B^+}(\infty)$, is 
varied. These results correspond to bulk $\mbox{pH}  =  5$ 
and $\mbox{pK}_{B^+} \rightarrow -\infty$ so that counterions coming from the salt don't adsorb on the 
chains. 

Monomer density profiles (Fig. ~\ref{fig:cseffect_ph5_1dip}(a)) 
show a depletion zone near the substrate, which is an outcome of
using Dirichlet boundary conditions for the chain propagator at the substrate and a delta function
as the initial condition. These boundary conditions correspond to a
non-adsorbing and highly repulsive substrate. Also, the monomer density profiles approach
zero around $z/R_{go} = 6-10$, depending on the bulk salt concentration.
Furthermore, the maxima in the monomer density profiles decreases in magnitude with an increase in the
bulk salt concentration. In other words, the height of the brush as manifest in the total
extent of the chains increases with an increase in the
bulk salt concentration. This is an outcome of an increase in the local charge inside
the brush as the bulk salt concentration
increases, which is reflected in Fig. ~\ref{fig:cseffect_ph5_1dip}(c).

The monomer density profiles affect the dielectric function in a significant manner as shown in 
Fig. ~\ref{fig:cseffect_ph5_1dip}(b). 
First, it is worth noting that the dielectric function closely follows the monomer density and 
a dielectric decrement is observed inside the brush region. The dielectric decrement is an 
outcome of the fact that the acidic groups on the chains are less polar than 
the solvent ($p_{H^+} < p_{s}$). The actual magnitude of the local dielectric 
function depends on the local density, electric field, temperature and the 
dipole moment of ion-pairs. In this work, we have fixed the temperature at room temperature 
by choosing $l_{Bo} = 56$ nm. In the region far from the grafting plane where 
the monomer density is zero, a dielectric 
constant of $80$ is obtained, representing solvent rich areas. 
Inside the brush, the dielectric decrement is as high as $30$, and it is seen that the 
dielectric decrement 
decreases with an increase in the bulk salt concentration. The decrease in the dielectric decrement is a 
result of a decrease in the local monomer density 
due to an increase in the local charge, as shown in Fig. ~\ref{fig:cseffect_ph5_1dip}(c). 

From Fig. ~\ref{fig:cseffect_ph5_1dip} (c) it is found that the 
probability of finding a monomer in the charged state ($\beta^\star(z)$) 
is higher in the bulk solution and decreases inside the brush region followed by an increase 
in the depletion zone near the substrate. This is in agreement with the fact that 
the degree of ionization of polyelectrolytes decreases with an increase in monomer density\cite{kumar_condensation}. 
Also, comparing the results from 
the calculations for different bulk salt concentrations, it is clear that the 
magnitude of $\beta^\star(z)$ increases with an increase in the bulk salt concentration. 
In other words, extent of charging of the polyelectrolyte chains increases with an increase 
in the bulk salt concentrations. This is a result of the decrease in the electrostatic potential 
with an increase in the bulk salt concentration (Fig. ~\ref{fig:cseffect_ph5_1dip} (d)), which leads to a 
shift in $\mbox{pK}_{H^+}$ (Eq. ~\ref{eq:charging_para}). 
Similar effects of the salt concentration on the charge regulation 
are observed in recent theoretical calculations\cite{szleifer_prl} 
and experiments\cite{granick}.  

The non-monotonic behavior of the probability of finding a monomer in 
the charged state is closely related to the non-monotonic behavior of the 
electrostatic potential as shown in Fig. ~\ref{fig:cseffect_ph5_1dip} (d).
It is observed that the electrostatic potential 
is negative everywhere for the polyacidic brush 
and approaches zero as far as $z/R_{go} = 50$ for the cases where the 
monomer density approaches zero at $z/R_{go} \sim 6-10$ 
(cf. Fig. ~\ref{fig:cseffect_ph5_1dip} (a)). 
This long-range 
behavior of the electrostatic potential complicates the numerical investigation 
of the strongly charged polyelectrolyte brushes in the absence of salt. 
We haven't extended the study to salt-free strongly charged polyelectrolyte brushes 
precisely because of this reason. However, we expect the results to be qualitatively similar.

Inside the brush region (i.e., $z/R_{go}< 6-10$ depending on the bulk salt concentration), the 
electrostatic potential closely follows 
the monomer density profiles (compare 
Figs. ~\ref{fig:cseffect_ph5_1dip} (a) and ~\ref{fig:cseffect_ph5_1dip} (d)). 
The minima in the electrostatic potential corresponds to the maxima in the monomer density profiles and 
inside the depletion zone, the magnitude of the electrostatic potential decreases due to a 
decrease in the monomer number density. 
An increase in the bulk salt concentration leads to an increase in the local 
degree of charging ($\beta^\star(z)$) and the electrostatic potential 
also increases in magnitude, as seen in Fig. ~\ref{fig:cseffect_ph5_1dip} (d). 

\begin{figure}[ht]
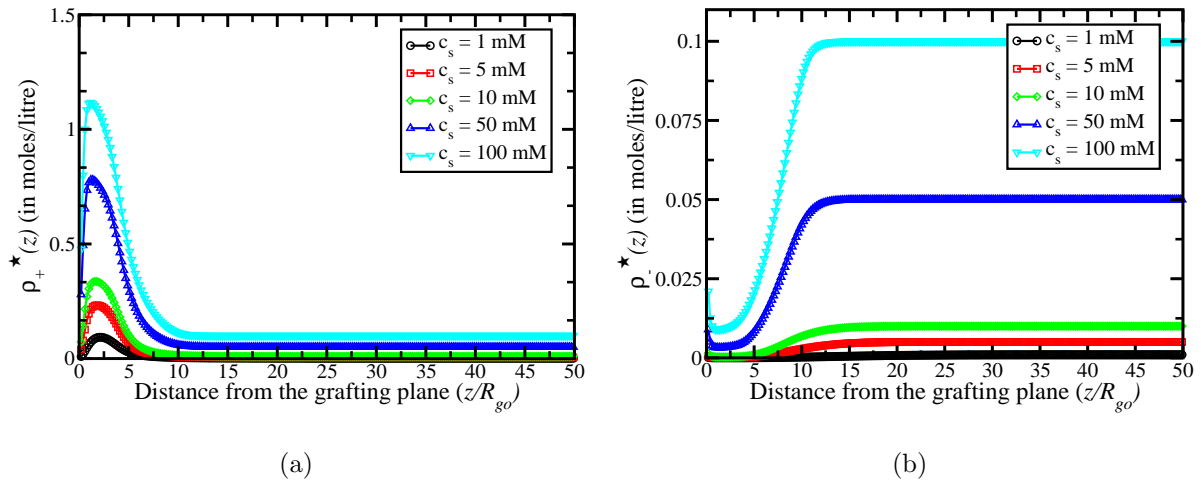
  \centering
\vspace{0.2in}
\subfigure[]{\includegraphics[width=3in]{cs_effect_count_wsolv1.eps}}
\hspace{0.1in}\subfigure[]{\includegraphics[width=3in]{cs_effect_coion_wsolv1.eps}
}
\caption{Distributions of small ions at $\mbox{pH} = 5$ and different salt concentrations in the case 
where counterions from the salt don't bind on to the chains. 
Figs. (a) and (b) show the total counterion and co-ion density 
profiles, respectively. The parameters used for producing these figures are the same as 
in Fig. ~\ref{fig:cseffect_ph5_1dip} and the data is cut at $z/R_{go} = 50$.}
\label{fig:ions_cseffect_ph5_1dip}
\end{figure}
  
\subsubsection{Counterion and coion distribution}
Figs. ~\ref{fig:ions_cseffect_ph5_1dip}(a) and ~\ref{fig:ions_cseffect_ph5_1dip}(b) show the 
density profiles for the ``free'' counterions (i.e., positively charged $H^+$ and $B^+$ ions) and 
coions (i.e., negatively charged $A^-$ ions), respectively, for different concentrations 
of the salt as set by the bulk solution. 
Physically, electrostatic attraction between the charged monomers and the counterions 
are responsible for an increase in the density of the counterions inside the brush region in comparison 
with the bulk.
Similarly, electrostatic repulsion between the charged monomers and co-ions
are the reason behind a lower number density of co-ions inside the brush region compared to the bulk 
solution as shown in Fig. ~\ref{fig:ions_cseffect_ph5_1dip}(b).

The magnitudes of the 
electrostatic attraction and repulsion for the monomer-counterion and 
monomer-co-ion pairs, respectively, are related to the electrostatic potential profiles 
shown in Fig. ~\ref{fig:cseffect_ph5_1dip}(d).
The non-monotonic behavior of the electrostatic potential manifests in the counterion and 
co-ion density profiles.
Furthermore, the strength of these
interactions is modulated by a change in the bulk salt concentration.
For example, an increase in the bulk salt concentration leads to an
increase in the degree of charging of monomers ($\beta^\star(z)$) and electrostatic 
potential ($\psi^\star(z)$), and increases 
the effect of electrostatic attractions and repulsions, as seen in Fig. ~\ref{fig:ions_cseffect_ph5_1dip}.

At this point, we comment on an important issue not considered in this work. 
We note that non-trivial effects such as the partitioning of small ions based on the 
ion solvation energy are not taken into account. However, we can infer 
the effects of solvation energy by the fact that solvation energy of small ions 
(within Born's theory\cite{intermolecular_forces}) is inversely proportional to the dielectric constant. 
This, in turn, means that the ions prefer regions of higher dielectric functions 
over regions of low dielectric functions. In the case of the co-ions inside a polyelectrolyte 
brush showing dielectric decrement, this effect 
works in synergy with the electrostatic repulsion between the monomers and the ions. 
In contrast, in the case of the counterions, the solvation effect 
counteracts the electrostatic attraction between the monomers and the coions. Depending on the 
relative strength of these counteracting effects, the counterion density 
may show a non-monotonic behavior 
when the solvation effects are taken into account. However, treatment of the solvation energy of charges 
on the chains complicates the analysis and this important direction of research 
is reserved for future work. 

\subsection{Effects of the competitive counterion adsorption}
\begin{figure}[ht]
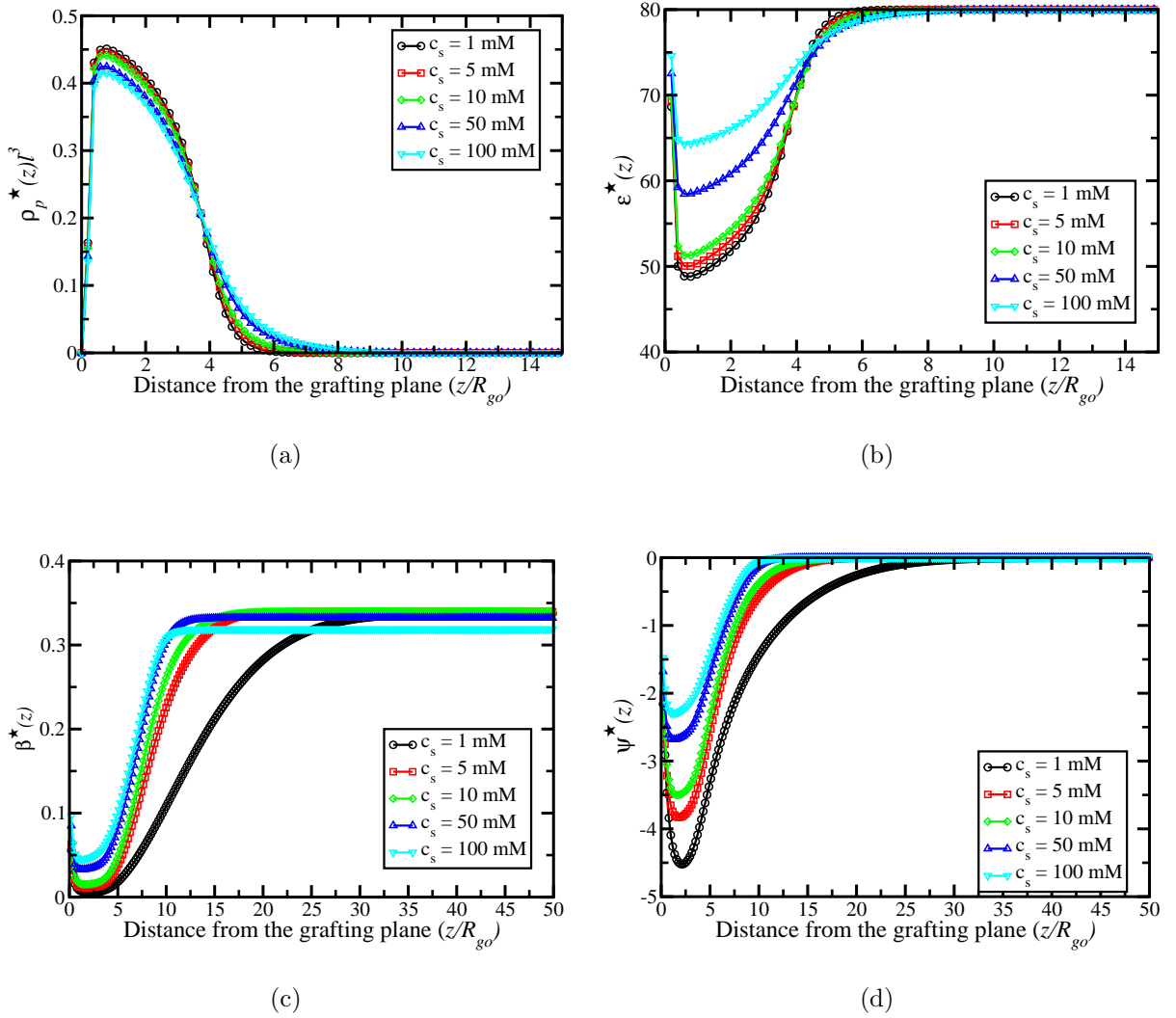
  \centering
\vspace{0.2in}
\subfigure[]{\includegraphics[width=3in]{cs_effect_monomer_wsolv2.eps}}
\hspace{0.1in}\subfigure[]{
\includegraphics[width=3in]{cs_effect_dielectric_wsolv2.eps}}\vspace{0.3in}\\
\subfigure[]{\includegraphics[width=3in]{cs_effect_ionization_wsolv2.eps}}
\hspace{0.1in}\subfigure[]{
\includegraphics[width=3in]{cs_effect_elec_potential_wsolv2.eps}}
\caption{Effect of the bulk salt concentrations on properties of the
polyacidic brush in the case where counterions from the salt can also bind for 
the bulk $\mbox{pH} = 5$ and $\mbox{pK}_{B^+} = 0.28$.
All the other parameters are the same as in Fig. ~\ref{fig:cseffect_ph5_1dip}. 
Figs. (a),(b),(c) and (d) correspond to
the monomer density ($\rho_p^\star(z)$), local dielectric
function ($\epsilon^\star(z)$), probability of finding a monomer in a charged state ($\beta^\star(z)$) and
the electrostatic potential ($\psi^\star(z)$), respectively. Like Fig. ~\ref{fig:cseffect_ph5_1dip}, 
the data is cut at $z/R_{go} = 15$ and $50$ to highlight the important
features for the Figs. (a)-(b) and (c)-(d), respectively.}
\label{fig:cseffect_ph5_2dip}
\end{figure}

\begin{figure}[ht]
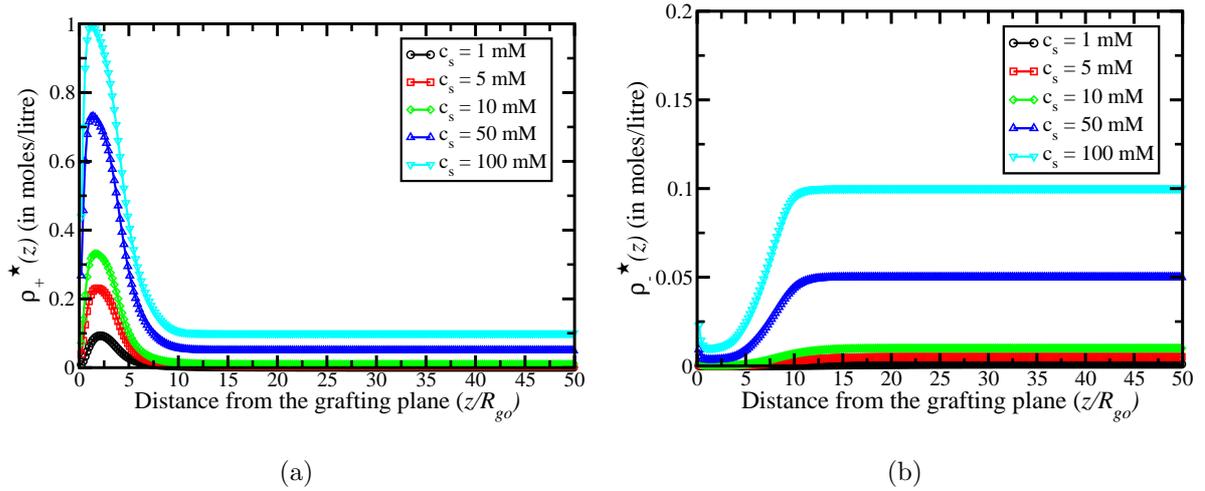
  \centering
\vspace{0.2in}
\subfigure[]{\includegraphics[width=3in]{cs_effect_count_wsolv2.eps}}
\hspace{0.1in}\subfigure[]{\includegraphics[width=3in]{cs_effect_coion_wsolv2.eps}}
\caption{Distributions of the counterions (a) and co-ions (b) for $\mbox{pH} = 5, 
\mbox{pK}_{B^+} = 0.28$ and different salt concentrations in the bulk. 
Monomer densities and other properties of these brushes are shown in Fig. ~\ref{fig:cseffect_ph5_2dip}.}
\label{fig:ions_cseffect_ph5_2dip}
\end{figure}

Fig. ~\ref{fig:cseffect_ph5_2dip} shows different characteristics of 
the poly(acid) brushes for $\mbox{pK}_{B^+} = 0.28$ so that counterions from the added salt  
can also adsorb on the polyelectrolyte chains.
From Figs. ~\ref{fig:cseffect_ph5_1dip} and ~\ref{fig:cseffect_ph5_2dip}, it is found 
that the monomer density profiles and the dielectric function
profiles are qualitatively similar in the two cases of counterion 
adsorption. However, there are quantitative differences and in particular, the 
charge regulation is affected in a significant manner as seen in Fig. ~\ref{fig:cseffect_ph5_2dip}(c).
For example, with an increase in the bulk salt concentration, $\beta^\star(z)$ in the bulk solution,
first increases (for $c_s \leq 10$ mM) and then decreases (for $c_s = 50, 100$ mM).
However, inside the brush region, $\beta^\star(z)$ increases with an increase 
in the salt concentration, as seen in Fig. ~\ref{fig:cseffect_ph5_1dip}(c) also.
The increase in the $\beta^\star(z)$ with an increase in the salt concentration
should be interpretted as a screening effect, which modifies the 
electrostatic potential (cf. Fig. ~\ref{fig:cseffect_ph5_2dip}(d)) and in turn,
the extent of counterion adsorption. This effect depends on the concentration 
of ``free'' ions (see Figs. ~\ref{fig:ions_cseffect_ph5_2dip}(a) 
and ~\ref{fig:ions_cseffect_ph5_2dip}(b)). 
However, in the case where counterions from the salt can also adsorb, 
Eqs. ~\ref{eq:binding_bulk1} and ~\ref{eq:binding_bulk2}, which are strictly 
valid at zero monomer density, reveal that  
an increase in the salt concentration at a fixed $\mbox{pH}$ leads to the 
adsorption of more counterions from the salt due to enhanced dissociation of protons. 
This effect is observed in Fig. ~\ref{fig:cseffect_ph5_2dip}(c) for finite 
monomer density at $c_s = 50, 100$ mM, 
where the bulk value of $\beta^\star(z)$ decreases with an increase in the bulk salt concentration. 

In other words, an increase in the bulk salt concentrations causes the counterion adsorption 
equilibrium to shift toward an increase in the adsorption of the counterions from the salt at 
the expense of dissociation of protons. Due to the fact that 
ion-pairs formed due to the adsorption of counterions from the salt are more polar than 
the native acid groups, the dielectric function is also affected by the shift in the counterion 
adsorption equilibrium. This can be observed by comparing Figs. 
~\ref{fig:cseffect_ph5_1dip}(b) and ~\ref{fig:cseffect_ph5_2dip}(b). For example, 
at $c_s = 100$ mM the lowest local dielectric function inside the brush is $~65$ for $\mbox{pK}_{B^+} = 0.28$ 
in comparison to the value of $\sim 50$ for $\mbox{pK}_{B^+} \rightarrow -\infty$. 
In the next section, we demonstrate that the shift in the 
counterion adsorption equilibrium from protons toward salt ions can also lead to dielectric 
increment just by modulating $\mbox{pH}$ of the bulk solution.  

\subsection{Effects of the pH of the bulk solution}
\begin{figure}[ht]
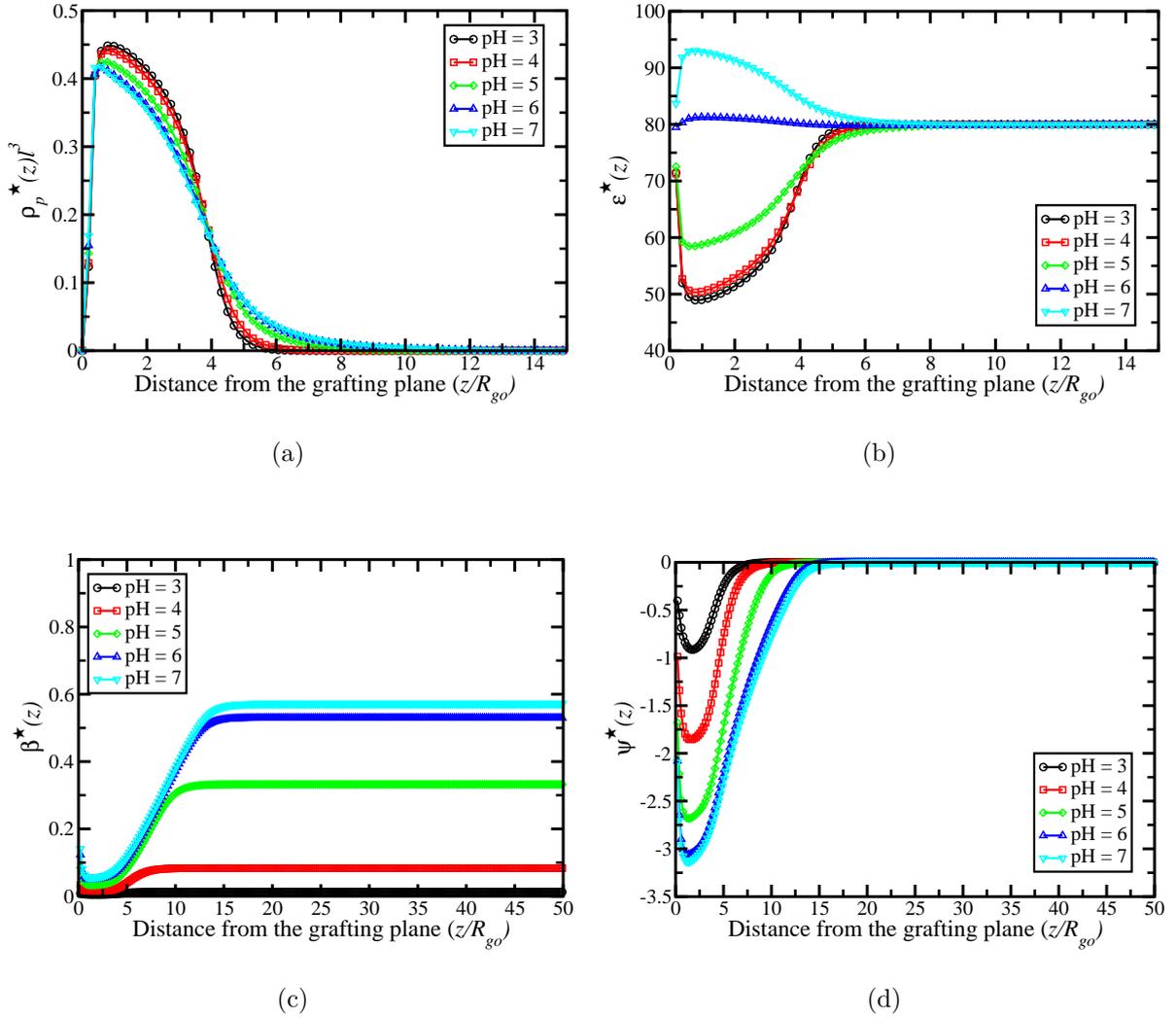
  \centering
\vspace{0.2in}
\subfigure[]{\includegraphics[width=3in]{ph_effect_monomer_wsolv.eps}}
\hspace{0.1in}\subfigure[]{\includegraphics[width=3in]{ph_effect_dielectric_wsolv.eps}}
\vspace{0.3in}\\
\subfigure[]{\includegraphics[width=3in]{ph_effect_ionization_wsolv.eps}}
\hspace{0.1in}\subfigure[]
{\includegraphics[width=3in]{ph_effect_elec_potential_wsolv.eps}} \\
\caption{Effect of the bulk $\mbox{pH}$ on the properties of poly(acid) brushes 
at a bulk salt concentration $=50$ mM and $\mbox{pK}_{B^+} = 0.28$. 
All the other parameters are the same as in Fig. ~\ref{fig:cseffect_ph5_1dip}.
Figs. (a),(b),(c) and (d) correspond to
the monomer density ($\rho_p^\star(z)$), local dielectric
function ($\epsilon^\star(z)$), probability of finding a monomer in a charged state ($\beta^\star(z)$) and
the electrostatic potential ($\psi^\star(z)$), respectively. Like Fig. ~\ref{fig:cseffect_ph5_1dip},
the data is cut at $z/R_{go} = 15$ and $50$ to highlight the important 
features for the Figs. (a)-(b) and (c)-(d), respectively.}
\label{fig:pheffect_csp05_2dip}
\end{figure}

It is well-known that an increase in pH of the bulk solution leads to an increase in
the charge on poly(acid) chains. This, in turn, means that there are
more sites for the adsorption of counterions from the salt. In
Fig. ~\ref{fig:pheffect_csp05_2dip}, we present the results
of calculations where dielectric increment is observed due to the replacement of less polar
acidic groups by more polar ion-pairs.

From Figs. ~\ref{fig:pheffect_csp05_2dip}(a) and ~\ref{fig:pheffect_csp05_2dip}(c), it
is clear that both the height of poly(acid) brushes and local charge 
increase with an increase in $\mbox{pH}$ of the bulk solution. Furthermore, the
electrostatic potential increases in magnitude as the bulk $\mbox{pH}$ increases.
However, the dielectric function inside the brush region is greater than that of the bulk
for $\mbox{pH} = 6,7$  in contrast to being lower for the other $\mbox{pH}$ values. 
As mentioned earlier, this is a direct outcome of salt ion adsorption onto the chains.

\begin{figure}[ht]
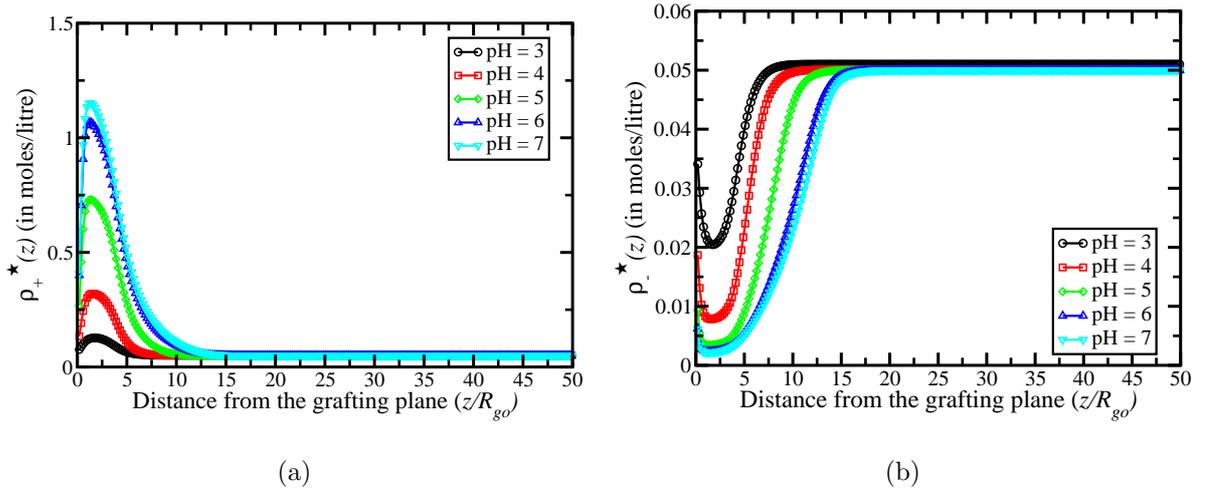
  \centering
\vspace{0.2in}
\subfigure[]{\includegraphics[width=3in]{ph_effect_count_wsolv.eps}}
\hspace{0.1in}\subfigure[]{\includegraphics[width=3in]{ph_effect_coion_wsolv.eps}}
\caption{Ion distributions for different $\mbox{pH}$ values in the bulk at a 
salt concentration $=50$ mM in the case 
where counterions from the salt also bind ($\mbox{pK}_{B^+} = 0.28$). 
Figs. (a) and (b) represents the counterion and co-ion density profiles, respectively.}
\label{fig:ions_pheffect_csp05_2dip}
\end{figure}

An increase in the local charge with an increase in the bulk $\mbox{pH}$ 
affects the counterion and co-ion density profiles, as shown in
Fig. ~\ref{fig:ions_pheffect_csp05_2dip}. An increase in the local charge leads to
an increase in the electrostatic potential, which manifests as an
increase in the counterion density profiles and as a decrease in the co-ion density profiles.

\subsection{Importance of non-linear effects}
\begin{figure}[ht]
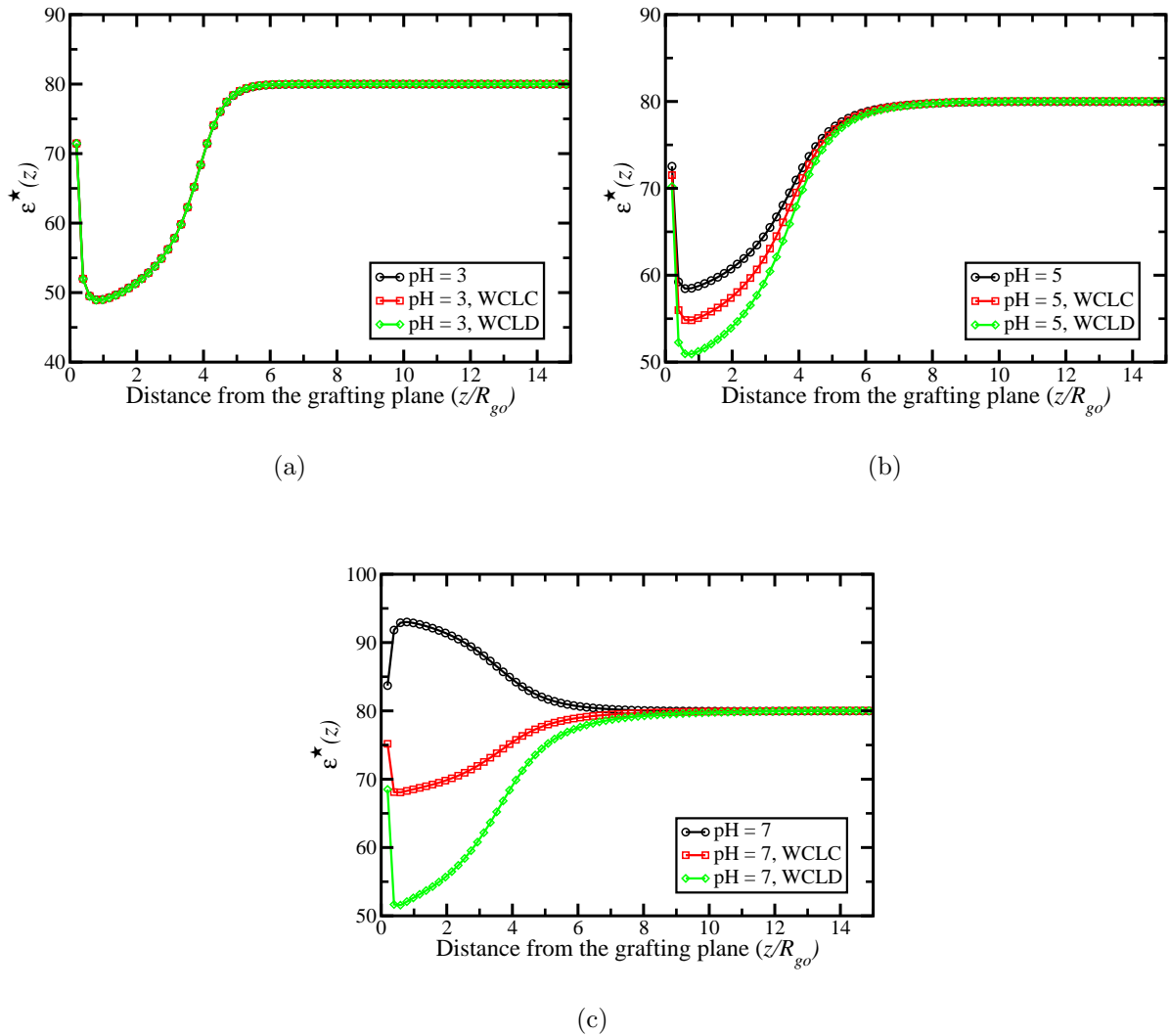
  \centering
\vspace{0.2in}
\subfigure[]{\includegraphics[width=3in]{compare_ph3_linear_wsolv.eps}}
\hspace{0.1in}\subfigure[]{\includegraphics[width=3in]{compare_ph5_linear_wsolv.eps}}
\vspace{0.3in}\\
\subfigure[]{\includegraphics[width=3in]{compare_ph7_linear_wsolv.eps}}
\caption{Significance of the non-linear effects on the dielectric function 
is demonstrated here. WCLC and WCLD represent two different 
ways of computing the dielectric function in the weak coupling limit. 
For the details, see the main text. The other plots for different 
pH values are the same as in Fig. ~\ref{fig:pheffect_csp05_2dip}(b). 
}
\label{fig:compare_linearmix}
\end{figure}

To highlight the importance of non-linear effects such as the dependence of the dielectric function 
on the local field and charge, we have compared our results with two 
approaches based on the assumption that the weak-coupling limit is valid 
independent of the coupling strength, 
determined by $p_{\gamma'}|\nabla_\mathbf{r}\psi^\star(\mathbf{r})|$. 
In particular, we have 
compared (Fig. ~\ref{fig:compare_linearmix}) the dielectric function predicted 
using the non-linear theory described herein 
with the behaviors expected from 
linear mixing rule for the dielectric function, which is shown to be 
valid in the weak coupling limit (cf. Eq. ~\ref{eq:bjerrum_weak}).
Also, to highlight the importance of non-linear effects on the charge 
regulation and, in turn, on the dielectric function, we have compared the results 
obtained by calculating $\beta_{H^+}$ and $\beta_{B^+}$ using either Eq. ~\ref{eq:charging_para}
(from the full non-linear theory) or 
Eqs. ~\ref{eq:binding_bulk1} - ~\ref{eq:binding_bulk2}, strictly valid in the limit of 
zero monomer density. These two approaches for the 
computation of dielectric function based on different estimates for $\beta_{H^+}$ and $\beta_{B^+}$ 
in the weak coupling limit are termed as 
WCLC (weak coupling limit with charge regulation) and WCLD (weak coupling limit in dilute regime), respectively. 
Note that the computations of the dielectric function in the WCLC and WCLD methods are done 
\textit{after} solving the SCFT equations for the full non-linear theory. In case, the non-linear effects 
are insignificant, the results obtained from the three methods (i.e., WCLC,WCLD and the full non-linear theory) 
should be identical. 
 
As seen from Fig. ~\ref{fig:compare_linearmix}, it is clear that the non-linear effects 
are not important when the degree of dissociation is low such as when pH$=3$ 
(cf. Figs. ~\ref{fig:compare_linearmix}(a) and ~\ref{fig:pheffect_csp05_2dip}(c)), as expected. 
In this case, non-linear effects on the charge regulation and, in turn, on the dielectric 
function are insignificant. In contrast, at a moderate degree of dissociation 
(e.g., for pH $=5$ in Fig. ~\ref{fig:compare_linearmix}(b)), there 
are noticeable differences between the dielectric functions predicted by 
the different approaches. In particular, comparing the dielectric functions 
obtained from the WCLC and WCLD methods in Fig. ~\ref{fig:compare_linearmix}(b), 
it is evident that the non-linear effects 
on charge regulation (arising from finite monomer density, cf. Eq. ~\ref{eq:charging_para}) 
play an important role. Furthermore, additional contributions arising from 
non-linear dependence of the dielectric function on the local field also play a significant 
role. These effects lead to a maximum difference of $\sim 5$ in the dielectric function, as seen 
in Fig. ~\ref{fig:compare_linearmix}(b). The non-linear effects 
are the most significant in cases where the charged groups are fully dissociated such as 
when the brush is in a solution at pH $= 7$, which is shown in Fig. ~\ref{fig:compare_linearmix}(c). 
In this case, the non-linear theory 
predicts a dielectric increment inside the brush, which is in striking contrast to 
the dielectric decrement predicted by the WCLC and WCLD methods. 
The origin of this discrepancy lies in the neglect of the dependence of 
local dipolar density ($p_{\gamma'}^\star(z)$) 
on the local field in the WCLC and WCLD methods. The dipolar density 
increases with an increase in the coupling strength 
($p_{\gamma'}|\nabla_\mathbf{r}\psi^\star(\mathbf{r})|$) as seen from Eq. ~\ref{eq:pstar}. 
However, the non-linear effects are not significant far from the brush regime where 
$p_{\gamma'}|\nabla_\mathbf{r}\psi^\star(\mathbf{r})| \rightarrow 0$
and the results obtained using the three different methods are the same, as expected. 

\section{Conclusions} \label{sec:conclusions}
We have studied planar polyelectrolyte brushes (made of end-tethered poly(acid) chains) in equilibrium with 
an electrolyte solution using field theory. We have focused 
on the quantitative description of the charge regulation and 
local dielectric function. Although the theory is quite general, in this 
work we have studied the effects of salt concentration and 
$\mbox{pH}$ of the bulk assuming a laterally homogeneous brush. In addition, the effects of 
competitive counterion adsorption 
are studied by allowing the salt ions to form ion-pairs by 
adsorption onto the charged monomers. 

The dipole moment of the ion-pairs is shown to significantly affect the dielectric function. 
For the poly(acid) chains bearing groups less polar than the solvent, the dielectric function 
inside the brush region is 
predicted to be lower than the bulk solution in the absence of adsorption          
of any salt ions. However, the formation of ion-pairs, generally, more polar than the 
solvent molecules, is shown to increase the dielectric function inside the brush 
in comparison to the bulk solution. Comparison of the theory taking into 
account dependence of the dielectric function on the local 
field and charge regulation (non-linear effects) with other approaches treating dielectric function 
by the linear mixing rules reveals that the non-linear effects are significant 
and must be taken into account in order to predict qualitatively correct 
behavior (such as the dielectric increment inside the brush) in charged systems.   
Furthermore, it is shown that local charge inside the brush region shows non-monotonic 
behavior. The local charge inside the brush region
increases with an increase in the bulk salt concentration. 
However, the counterion adsorption equilibrium shifts from 
protons to the salt ions with an increase in the salt concentration. 
Counterion and co-ion densities are predicted to be higher and lower, respectively, 
in comparison with their bulk value.  

We also comment on an important assumption used in the theory. 
We have ignored the effects of ``induced'' dipole moments, 
which have been shown to enhance net dipole moment\cite{onsager_moments}.  
Ignoring this induction effect is a reasonable first step before developing a 
more comprehensive theory to treat the dielectric response of flexible macromolecules. 
In the future, we plan to overcome this limitation of the theory by extending the field 
theoretical method developed here.  

Finally, it is understood that the dielectric function has an intricate relation with the
solvation\cite{intermolecular_forces,ninham_work1,ninham_work2,onuki_kitamura,wang_solvation} of charges.
Classic work by Born\cite{intermolecular_forces} directly relates
the solubility of charged molecules into solvents of different dielectric constants and predicts the partitioning
of salt ions based on their valency and radii. In the literature it has been shown that the solvation of
ions can lead to non-trivial and counter-intutive results. For example, salt induced
stabilization of the ordered morphologies in the mixtures of small
molecules\cite{sadakane_1,sadakane_2,sadakane_3,sadakane_4}
is shown to be an outcome of the
asymmetric\cite{nabutovskii_1,nabutovskii_2,onuki_kitamura} solvation of the cations and anions.
The theoretical treatment presented here employs the saddle-point approximation for 
point ions and dipoles. Thus, the theory in the current form is unable to capture effects such as 
asymmetric solvation of ions based on their radii. However, the theory can be extended to 
finite size ions by using a soft primitive model for charges\cite{wang_solvation2,hansen,rob_interfacial}. 
We plan to address this issue in a future publication. 

\section*{ACKNOWLEDGMENTS}
\setcounter {equation} {0} \label{acknowledgement}
This research used resources of the Oak Ridge Leadership Computing
Facility at the Oak Ridge National Laboratory, which is supported by the
Office of Science of the U.S. Department of Energy under Contract No.
DE-AC05-00OR22725. This research was conducted at the Center for Nanophase Materials Sciences, which is sponsored 
at Oak Ridge National Laboratory by the Scientific User Facilities Division, 
Office of Basic Energy Sciences, U.S. Department of Energy. 
SMKII acknowledges funding from the National Science Foundation (Grant
Nos. $0840249$ and $1133320$) which supports experimental efforts at the
University of Tennessee that in part motivated this study.

\renewcommand{\theequation}{A-\arabic{equation}}
  \setcounter{equation}{0}  
  \section*{APPENDIX A : Counterion adsorption, dipolar interactions and the partition function for polyelectrolyte brushes} \label{app:A}

Here we present the partition function for the polyelectrolyte brushes in the presence of polar 
solvent molecules and small monovalent ions. Special attention is paid to the counterion adsorption and 
dipolar interactions. We start from a molecular description taking into account the 
dipolar interactions along with other short range interactions (dispersion and 
van der Waals). In order to cast theory in a field theoretical language, we rewrite the 
partition function in a form suitable for mathematical transformations. 
In this work, we consider polyelectrolyte brushes made of $n$ mono-disperse end-tethered chains, each 
containing $N$ Kuhn segments. 

Following Edward's work\cite{edwardsbook}, we represent a polyelectrolyte chain in the brush 
by a continuous curve $\mathbf{R}(t)$ of length $Nl$, $l$ being the Kuhn segment length and $t$
is the arc variable representing any point on the curve lying in the range $(0,N)$. To keep 
track of different chains, subscript $\alpha$ is used so that $t_{\alpha}$
represents the contour variable along the backbone of $\alpha^{th}$ chain.
Also, the position vector for a particular segment,$t_{\alpha}$, is written as 
 $\mathbf{R}_{\alpha}(t_{\alpha})$. 

In the following, subscripts $p,s$ and $\gamma$ are used to
represent monomers, solvent molecules and the small ions, respectively.
Three different kinds of small ions are considered here and unless specified, 
$\gamma = H^+, B^+$ and $A^-$ represents
protons, cations from the salt and anions, respectively, as described in the main text. 
In this work, we study negatively charged (or polyacidic chains) chains and specificity of 
the cations is taken into account to study the effects of different binding energies of the cations.
For the treatment shown below, we assume that \textit{local} incompressiblity condition is satisfied. 
Treating all the small ions as point-like, the incompressiblity condition allows us to write
the total volume as $\Omega = n N/\rho_{po} + n_s/\rho_{so}$ where $\rho_{po}$ and $\rho_{so}$ are
the bulk monomer and solvent densities, respectively.
Also, we use the notation $l^{3} \equiv 1/\rho_{po}$ and $\mathbf{r}_k$ represents the
position vector of $k^{th}$ small molecule like solvent molecules and small ions.

Furthermore, counterion adsorption on the polyelectrolyte chains is taken into account using 
a two-state model. Segments along the chains can be either in charged or in uncharged state. 
To describe the two states, another arc length variable, $\theta_\alpha(t_\alpha)$, is introduced, 
which enumerates the state of charging of the segment,$t_{\alpha}$ on $\alpha^{th}$ chain.
For the analysis here, $\theta_\alpha(t_\alpha) = 0$ means $t_\alpha$ is a neutral site 
and $\theta_\alpha(t_\alpha) = 1$ represents a fully charged site along the backbone.
Like the average over all of the possible conformations in the theories of neutral polymers, we 
need to average over all of the possible charge distributions along the chains. We 
represent the average over $\theta_\alpha(t_\alpha)$ by symbol 
$\sum_{\left\{\theta_\alpha\right\}}\left < (\cdots) \right >$, which expicitly means
\begin{eqnarray}
\sum_{\left\{\theta_\alpha\right\}}\left < (\cdots) \right >  &=& \int_0^N \prod_{\alpha=1}^n dt_\alpha \int D\left[\theta_\alpha(t_\alpha)\right](\cdots)P\left[\theta_\alpha(t_\alpha)\right]\Upsilon\left[\theta_\alpha(t_\alpha)\right].
\end{eqnarray}
Here, $P\left[\theta_\alpha\right]$ is the probability distribution function for the variable $\theta_\alpha$, 
which must satisfy the relation $\int D\left[\theta_\alpha(t_\alpha)\right] P\left[\theta_\alpha(t_\alpha)\right] = 1$. 
Also, $\Upsilon\left[\theta_\alpha(t_\alpha)\right]$ is the number of indistinguishable 
ways in which $\theta_\alpha$ can be distributed among $nN$ sites for a fixed number of charged sites. 
$\Upsilon\left[\theta_\alpha(t_\alpha)\right]$ takes into account the entropy of distribution of charged sites. 

Like the segments, the counterions are also divided into two sets. 
One set of counterions is ``free'' to explore the whole space and has translational 
degrees of freedom. The other set is ``adsorbed'' on the backbone (ion-pairs) and 
behave as electric dipoles. Number of counterions in ``free'' and ``adsorbed'' 
states are taken to be $n_\gamma^f$ and $n_\gamma^a$, respectively, for 
$\gamma = H^+, B^+$. In this work, we treat the ion-pairs as \textit{point} electric dipoles for the 
development of an understanding of dielectric function. 
In the following, the dipole moment along the $\alpha^{th}$ chain backbone is 
written as by vector, $\mathbf{p}_{\alpha}(t_\alpha)$. 
Similarly, $\mathbf{p}_{k}$ represents the dipole moment of the $k^{th}$ solvent 
molecule. 

Using the notations described above, the 
partition function ($Z$) for a polyelectrolyte brush can be written as
\begin{eqnarray}
       Z & = & \int \prod_{\alpha=1}^n D[\mathbf{R}_\alpha] \sum_{\left\{\theta_\alpha(t_\alpha)\right\}}\left < \int \prod_{\alpha=1}^n\prod_{t_\alpha=0}^{N} d\mathbf{p}_{\alpha}(t_\alpha) 
\int \frac{1}{\prod_{\gamma'}n_{\gamma'}^f!n_{A^-}!n_s!}\prod_{j=1}^{n_{\gamma'} + n_s + n_{A^-}} d\mathbf{r}_{j} \int \prod_{k=1}^{n_s}d\mathbf{p}_{k} \right . \nonumber \\
&& \exp \left [-H_0\left\{\mathbf{R}_\alpha\right\}   - H_w\left\{\mathbf{R}_\alpha,\mathbf{R}_{\alpha'}\right\} 
- H_{cp}\left\{\mathbf{R}_\alpha,\mathbf{p}_{\alpha},\mathbf{r}_j\right\} 
- H_{pp}\left\{\mathbf{R}_\alpha,\mathbf{p}_{\alpha},\mathbf{R}_{\alpha'},\mathbf{p}_{\alpha'}\right\}\right . \nonumber \\
&& \left .  \left .  - H_{cc}\left\{\mathbf{r}_j,\mathbf{r}_j'
\right\} -E\left\{\theta_\alpha\right\}\right ] 
\prod_{\mathbf{r}}\mathbf{\delta}\left[\frac{\hat{\rho}_{p}(\mathbf{r})}{\rho_{po}} + \frac{\hat{\rho}_{s}(\mathbf{r})}{\rho_{so}} - 1\right]\right > \label{eq:parti_sing}
\end{eqnarray}
where $\gamma' = H^+,B^+$.

The Hamiltonian in Eq. ~\ref{eq:parti_sing} is written by taking into account
the contributions coming from the chain connectivity (given by $H_0$ in Eq. ~\ref{eq:connectivity} below), 
the short ranged dispersion interactions (represented by $H_w$ in Eq. ~\ref{eq:dispersion}) 
and the long range electrostatic interactions between the charged species 
(written as $H_{pp}, H_{cp}$ and $H_{cc}$ above, which correspond to the dipole-dipole, charge-dipole 
and charge-charge interactions, respectively). For convenience in writing, in the following we have 
suppressed the explicit functional dependence of $H_{0},H_w, H_{pp}, H_{cp}$ and $H_{cc}$.

Explicitly, contributions from the chain connectivity are given by:
\begin{eqnarray}
      H_0 &=& \frac {3}{2 l^2}\sum_{\alpha=1}^n \int_{0}^{N} dt_\alpha \left(\frac{\partial \mathbf{R}_\alpha(t_\alpha)}{\partial t_\alpha} \right )^{2} \label{eq:connectivity}
\end{eqnarray}
which is the well-known Wiener measure for a
flexible polymer chain. Furthermore, $H_{w}\left\{\mathbf{R}_\alpha,\mathbf{R}_{\alpha'}\right\}$ 
takes into account the energetic contributions coming from short range dispersion interactions between segments 
of chains indexed as $\alpha$ and $\alpha'$ located at $\mathbf{R}_\alpha$ and $\mathbf{R}_{\alpha'}$, respectively. 
Following Edwards's formulation\cite{edwardsbook} for a flexible chain, we model these interactions by 
delta functional/point 
interactions, which allows us to write 
\begin{eqnarray}
H_w &=& \frac{1}{2}\int d\mathbf{r} 
\left[w_{pp}\hat{\rho}_{p}^2(\mathbf{r}) + w_{ss}\hat{\rho}_{s}^2(\mathbf{r})  + 2 w_{ps} \hat{\rho}_{s}(\mathbf{r})
\hat{\rho}_{s}(\mathbf{r})\right] \label{eq:dispersion}
\end{eqnarray}
Here, $w_{pp}, w_{ss}$ and $w_{ps}$ are the well-known excluded volume parameters describing the 
strength of interactions between $p-p, s-s$ and $p-s$ pairs, respectively. Also, 
$\hat{\rho}_{p}(\mathbf{r})$ and $\hat{\rho}_{s}(\mathbf{r})$ represent the microscopic number 
density of the monomers and the solvent molecules, respectively, at a 
certain location $\mathbf{r}$ defined as
        \begin{eqnarray}
\hat{\rho}_{p}(\mathbf{r})  &=& \sum_{\alpha=1}^n \int_{0}^{N} dt_\alpha \, \delta \left[\mathbf{r}-\mathbf{R}_\alpha(t_\alpha)\right] \\
\hat{\rho}_{s}(\mathbf{r})  &=& \sum_{k=1}^{n_s} \, \delta \left[\mathbf{r}-\mathbf{r}_k\right]
\end{eqnarray}

Electrostatic contributions to the Hamiltonian arising from the dipole-dipole interactions can be written 
by considering \textit{finite/real} dipoles first and then systematically approaching the limit of 
\textit{point} dipoles so that the dipole length goes to zero and the charge increases keeping 
the dipole moment finite. For \textit{real} dipoles, we can explicitly 
add the Coulomb 
interaction energies between the charges at the ends of dipoles on the $\alpha^{th}$ polyelectrolyte chain 
and solvent molecules. The magnitude of the charges (in units of electronic charge, $e$) 
at the end of dipoles on the chains and solvent molecules are written as 
$Z_{pk}$ and $Z_{sk}$, respectively so that $k=\pm$ and $Z_{p+} = - Z_{p-}, Z_{s+} = - Z_{s-}$. 
For the dipoles located on the 
polyelectrolyte chains, we consider their centers to lie on 
the chain backbone, and the 
dipolar axes along the vector $\mathbf{p}_\alpha(t_\alpha)$. 
Summing over all the dipoles, we 
obtain the dipole-dipole interaction energy as 

\begin{eqnarray}
H_{pp} &=& \frac{l_{Bo}}{2}\int d\mathbf{r}\int d\mathbf{r}' \int d\mathbf{p}\int d\mathbf{p}'\left[ \left \{ \frac{Z_{p+}^2}{|\mathbf{r} - \mathbf{r}' + 0.5\mathbf{p} -0.5\mathbf{p}'|} + \frac{Z_{p-}^2}{|\mathbf{r} - \mathbf{r}' - 0.5\mathbf{p} + 0.5\mathbf{p}'|} \right . \right . \nonumber \\
&& \left . + \frac{Z_{p+}Z_{p-}}{|\mathbf{r} 
- \mathbf{r}' + 0.5\mathbf{p} + 0.5\mathbf{p}'|} +  \frac{Z_{p+}Z_{p-}}{|\mathbf{r} - \mathbf{r}' - 0.5\mathbf{p} 
- 0.5\mathbf{p}'|}\right \}
\bar{\rho}_{p}(\mathbf{r},\mathbf{p} )\bar{\rho}_{p}(\mathbf{r}',\mathbf{p}') \nonumber \\
&& + \left\{ \frac{Z_{s+}^2}{|\mathbf{r} - \mathbf{r}' + 0.5\mathbf{p} -0.5\mathbf{p}'|} + 
\frac{Z_{s-}^2}{|\mathbf{r} - \mathbf{r}' - 0.5\mathbf{p} + 0.5\mathbf{p}'|}  \right . \nonumber \\
&& \left . + \frac{Z_{s+}Z_{s-}}{|\mathbf{r}- \mathbf{r}' + 0.5\mathbf{p} + 0.5\mathbf{p}'|} +  
\frac{Z_{s+}Z_{s-}}{|\mathbf{r} - \mathbf{r}' - 0.5\mathbf{p} - 0.5\mathbf{p}'|} \right \}
\bar{\rho}_{s}(\mathbf{r},\mathbf{p} )\bar{\rho}_{s}(\mathbf{r}',\mathbf{p}') \nonumber \\
&& + 2\left\{ \frac{Z_{p+}Z_{s+}}{|\mathbf{r} - \mathbf{r}' + 0.5\mathbf{p} -0.5\mathbf{p}'|} +
\frac{Z_{p-}Z_{s-}}{|\mathbf{r} - \mathbf{r}' - 0.5\mathbf{p} + 0.5\mathbf{p}'|}  \right . \nonumber \\
&& \left . \left . + \frac{Z_{p+}Z_{s-}}{|\mathbf{r}- \mathbf{r}' + 0.5\mathbf{p} + 0.5\mathbf{p}'|} +
\frac{Z_{p-}Z_{s+}}{|\mathbf{r} - \mathbf{r}' - 0.5\mathbf{p} - 0.5\mathbf{p}'|} \right \}
\bar{\rho}_{p}(\mathbf{r},\mathbf{p} )\bar{\rho}_{s}(\mathbf{r}',\mathbf{p}') \right ]\label{eq:segelec}
\end{eqnarray}
where $l_{Bo} = e^2/\epsilon_o k_B T$ is the Bjerrum length in vacuum and $\epsilon_o$ is the 
permittivity of the vacuum. Also, $k_B T$ is the Boltzmann constant times the temperature. 
Also, in Eq. ~\ref{eq:segelec}, we have defined the microscopic number density of dipoles on the chains and solvent molecules by 
$\bar{\rho}_p(\mathbf{r},\mathbf{p})$ 
and $\bar{\rho}_s(\mathbf{r},\mathbf{p})$, respectively. 
Physically, these functions describe the number of dipoles with 
their centers at a certain location $\mathbf{r}$ with dipole moments given by $\mathbf{p}$. 
Formally, these are defined as
\begin{eqnarray}
    \bar{\rho}_{p}(\mathbf{r},\mathbf{p})  &=& \sum_{\alpha=1}^n \int_{0}^{N} dt_\alpha \, \delta \left[\mathbf{r}
-\mathbf{R}_\alpha(t_\alpha)\right]
\delta \left[\mathbf{p}-\mathbf{p}_\alpha(t_\alpha)\right]\left[1 - \theta_\alpha(t_\alpha)\right] \\
\bar{\rho}_{s}(\mathbf{r},\mathbf{p})  &=& \sum_{k=1}^{n_s} \, \delta \left[\mathbf{r}-\mathbf{r}_k\right]
\delta \left[\mathbf{p}-\mathbf{p}_k\right]
\end{eqnarray}
Repeating the same arguments as above, we can write the interaction energy between the charges and 
the dipoles as 
\begin{eqnarray}
H_{cp} &=&  l_{Bo}\int d\mathbf{r}\int d\mathbf{r}' \int d\mathbf{p}'\left[\sum_{\gamma} Z_\gamma \hat{\rho}_{\gamma}(\mathbf{r}) + Z_p \hat{\rho}_{pe}(\mathbf{r})\right]
\left[\frac{Z_{p+}\bar{\rho}_{p}(\mathbf{r}',\mathbf{p}' )}{|\mathbf{r} - \mathbf{r}' - 0.5\mathbf{p}'|} \right . \nonumber \\
&& \left. 
+ \frac{Z_{p-}\bar{\rho}_{p}(\mathbf{r}',\mathbf{p}' )}{|\mathbf{r} - \mathbf{r}' + 0.5\mathbf{p}'|} + 
\frac{Z_{s+}\bar{\rho}_{s}(\mathbf{r}',\mathbf{p}' )}{|\mathbf{r} - \mathbf{r}' - 0.5\mathbf{p}'|}
+ \frac{Z_{s-}\bar{\rho}_{s}(\mathbf{r}',\mathbf{p}' )}{|\mathbf{r} - \mathbf{r}' + 0.5\mathbf{p}'|} \right ] \label{eq:countseg}
\end{eqnarray}
where $\hat{\rho}_\gamma(\mathbf{r})$ represents the local microscopic densities for the point-like ions of type $\gamma$ at 
$\mathbf{r}$, defined as
        \begin{eqnarray}
     \hat{\rho}_{\gamma}(\mathbf{r}) &=&  \sum_{j=1}^{n_{\gamma}} \delta \left[\mathbf{r}-\mathbf{r}_j\right] \quad \mbox{for} \quad \gamma = H^+,B^+,A^-.
\end{eqnarray}
Furthermore, $\hat{\rho}_{pe}(\mathbf{r})$ is the contribution to the charge density 
coming from the polyelectrolyte chains, given by
\begin{eqnarray}
     \hat{\rho}_{pe}(\mathbf{r}) &=& \sum_{\alpha=1}^n \int_{0}^{N} dt_\alpha \, \delta \left[\mathbf{r} -\mathbf{R}_\alpha(t_\alpha)\right] \theta_\alpha(t_\alpha)
\end{eqnarray}

Electrostatic interaction energy among the small ions can be written using Coulomb's law as 
\begin{eqnarray}
H_{cc}  & = & \frac{l_{Bo}}{2}\int d\mathbf{r} \int d\mathbf{r}'\frac{\left[\sum_{\gamma} 
Z_\gamma\hat{\rho}_{\gamma}(\mathbf{r}) + Z_p \hat{\rho}_{pe}(\mathbf{r}) \right]\left[\sum_{\gamma} Z_\gamma \hat{\rho}_{\gamma}(\mathbf{r}') + Z_p \hat{\rho}_{pe}(\mathbf{r}') \right]}
{|\mathbf{r} - \mathbf{r}' |},
\label{eq:countcount}
\end{eqnarray}

The electrostatic terms, written for \textit{``finite''} electric dipoles, can be written in 
a simplified form, if we consider the limit of \textit{``point''} dipoles so that the length 
of the dipoles approaches zero. 
Also, as mentioned earlier, for the electric dipoles, $Z_{p+} = - Z_{p-} = Z_p$ 
and $Z_{s+} = -Z_{s-} = Z_s$.
Approaching the limits of infinitesimal dipole lengths in such a way that the dipole moments 
approach their finite values, we can use multipole expansion and write 

\begin{eqnarray}
H_e = H_{pp} + H_{cp} + H_{cc} &=& \frac{l_{Bo}}{2}\int d\mathbf{r}\int d\mathbf{r}' 
\frac{\left[\hat{\rho}_e(\mathbf{r}) - \nabla_{\mathbf{r}}.\hat{P}_{ave}(\mathbf{r}) \right]\left[\hat{\rho}_e(\mathbf{r}') - \nabla_{\mathbf{r}'}.\hat{P}_{ave}(\mathbf{r}')
\right]}{|\mathbf{r} - \mathbf{r}'|} \nonumber \\
&&\label{eq:final_particle_elec}
\end{eqnarray}
where  $\hat{\rho}_e(\mathbf{r}) = \sum_{\gamma} Z_\gamma \hat{\rho}_{\gamma}(\mathbf{r}) + Z_p \hat{\rho}_{pe}(\mathbf{r})$ 
is the local charge density. Also, $\hat{P}_{ave}(\mathbf{r}') 
= \int d\mathbf{p} \hat{P}(\mathbf{r},\mathbf{p})\mathbf{p}$, is a vectorial quantity so that 
$\hat{P}(\mathbf{r},\mathbf{p} ) = \bar{\rho}_p(\mathbf{r},\mathbf{p} ) + \bar{\rho}_s(\mathbf{r},\mathbf{p} )$ 
is local dipole density
at $\mathbf{r}$ with the dipole moment $\mathbf{p}$. Such an expression for the 
electrostatic contributions resulting from polarization was proposed by Marcus\cite{marcus} and 
Felderhof\cite{felderhof}. Furthermore, statistical mechanical studies 
have been carried out for small molecules\cite{coalson,orland_dipolar,wang08} 
using this expression for the electrostatics in the free energy. 

Note that the electrostatic terms written above depends on the arc length variable $\theta_\alpha(t_\alpha)$. 
This variable also determines the energetic contributions of counterion adsorption on the backbone, written as
$E\left\{\theta_\alpha\right\}$ in Eq. ~\ref{eq:parti_sing}. Noting that 
the dissociable groups on the chains have to dissociate first for the salt ions to 
adsorb, it is written as 
\begin{eqnarray}
E\left\{\theta_\alpha\right\} = (nN - n_{H^+}^a) \left[\mu_{COO^-}^o + \mu_{H^+}^o - \mu_{COOH}^o\right]
 + n_{B^+}^a \left[\mu_{COO^-B^+}^o - \mu_{COO^-}^o - \mu_{B^+}^o \right]
\end{eqnarray}
where $n_{\gamma}^a$ is the number of counterions of type $\gamma'=H^+,B^+$ adsorbed on the backbone. 
$\mu_j^o$ is the chemical potential (in units of $k_B T$) for species of type $j$ in infinitely dilute conditions.
The differences in the chemical potentials are related to the equilibrium constants 
of the corresponding reactions by the relations\cite{mcquarie}
\begin{eqnarray}
\mu_{COO^-}^o + \mu_{H^+}^o - \mu_{COOH}^o &=& 2.303 \mbox{pK}_{H^+} = -2.303 \log_{10}\mbox{K}_{H^+} \\
\mu_{COO^-}^o + \mu_{B^+}^o - \mu_{COO^-B^+}^o &=& 2.303 \mbox{pK}_{B^+} = -2.303 \log_{10}\mbox{K}_{B^+} 
\end{eqnarray}
Here, we have defined $\mbox{K}_{H^+}$ and $\mbox{K}_{B^+}$ as the equilibrium (dissociation) 
constants for the reactions 
\begin{eqnarray}
-COOH &\rightleftharpoons& -COO^- + H^+ \\
-COO^-B^+ &\rightleftharpoons& -COO^- + B^+
\end{eqnarray}
respectively. Such a model of counterion adsorption was developed in the classic paper by Harris 
and Rice\cite{harris_rice}.

\renewcommand{\theequation}{B-\arabic{equation}}
  \setcounter{equation}{0}  
  \section*{APPENDIX B : Field theory for polyelectrolyte brushes : annealed charge distribution} \label{app:B}

The probability distribution, $P$, needs to be determined self-consistently by the
minimization of the free energy and must
satisfy the relation $\int D\left[\theta_\alpha(t_\alpha)\right] P\left[\theta_\alpha(t_\alpha)\right] = 1$.
In this work, we take a variational \textit{ansatz} for $P$ and write it as
\begin{eqnarray}
P\left[\theta_\alpha(t_\alpha)\right] &=& \left(\sum_{\gamma'=H^+,B^+}\beta_{\gamma'}\right)
\delta\left[\theta_\alpha(t_\alpha)\right] + \left(1-\sum_{\gamma'=H^+,B^+}\beta_{\gamma'}\right)\delta\left[\theta_\alpha(t_\alpha) - 1\right] \label{eq:prob_dist}
\end{eqnarray}
so that $n_{\gamma'}^a = \beta_{\gamma'} n N$ for $\gamma' = H^+,B^+$.
Mathematically, $\beta_{H^+}$ and $\beta_{B^+}$ are the variational parameters, which will be determined by 
minimization of the free energy. Physically, $\beta_{H^+}$ and $\beta_{B^+}$ correspond to 
the fraction of sites on the chains occupied by $H^+$ and $B^+$, respectively. 
Furthermore, using Eq. ~\ref{eq:prob_dist} for the charge distribution
\begin{eqnarray}
\Upsilon\left[\theta_\alpha(t_\alpha)\right] &=& \frac{nN!}{n_{H^+}^a! n_{B^+}^a!(nN- n_{H^+}^a - n_{B^+}^a)!}
\end{eqnarray}
Such a distribution is called ``annealed'' distribution in the literature\cite{borukhov,shi_scft,wang_scft}. 
The use of Eq. ~\ref{eq:prob_dist} for the probability distribution the charged sites on the 
chain backones simplifies the subsequent analysis. 
Using Eq. ~\ref{eq:prob_dist}, we can write Eq. ~\ref{eq:parti_sing} in field theoretic form. 
We start from the electrostatic contribution to the partition function, which is given 
by Eq. ~\ref{eq:final_particle_elec}, and use the Hubbard-Statonovich 
transformation\cite{fredbook}

\begin{eqnarray}
\exp\left[-H_e\right] &=& \frac{1}{\Xi} \int D\left[\psi\right]\exp \left[-i\int d\mathbf{r}
\left\{\hat{\rho}_e(\mathbf{r}) - \nabla_{\mathbf{r}}.\hat{P}_{ave}(\mathbf{r})\right\} \psi(\mathbf{r}) +
\frac{1}{8\pi l_{Bo}}\int d\mathbf{r}\psi(\mathbf{r})\nabla_\mathbf{r}^2 \psi(\mathbf{r}))\right], \nonumber \\
&&
\end{eqnarray}
where
\begin{eqnarray}
\Xi &=& \int D\left[\psi\right]\exp \left[\frac{1}{8\pi
l_{Bo}}\int d\mathbf{r}\psi(\mathbf{r})\nabla_\mathbf{r}^2 \psi(\mathbf{r}))\right].
\end{eqnarray}
Using this transformation and Eq. ~\ref{eq:prob_dist}, we can integrate over 
the orientations of the dipoles analytically and evaluate the average over $\theta_\alpha$. 
Writing the magnitude of the dipole moments of solvent molecules as $p_s$ and ion-pairs 
as $p_{\gamma'}$ for $\gamma' = H^+,B^+$, the 
partition function given by Eq.~\ref{eq:parti_sing} becomes
\begin{eqnarray}
       Z & = & e^{-F_a/k_B T}\int \prod_{\alpha=1}^n D[\mathbf{R}_\alpha] 
\int \prod_{\gamma} \prod_{j=1}^{n_{\gamma}} 
d\mathbf{r}_{j} \prod_{k=1}^{n_{s}} d\mathbf{r}_{k}\frac{1}{\Xi} \int D\left[\psi\right]\exp \left [-H_0\left\{\mathbf{R}_\alpha\right\} 
- H_w\left\{\mathbf{R}_\alpha,\mathbf{R}_{\alpha'}\right \} \right .\nonumber \\
&&  - i\int d\mathbf{r} \sum_{\gamma}Z_\gamma \hat{\rho}_\gamma(\mathbf{r})\psi(\mathbf{r}) - i \int d\mathbf{r} \hat{\rho}_p(\mathbf{r})\psi_p(\mathbf{r}) +
\frac{1}{8\pi l_{Bo}}\int d\mathbf{r}\psi(\mathbf{r})\nabla_\mathbf{r}^2 \psi(\mathbf{r})) \nonumber \\
&& \left . + \int d\mathbf{r} \hat{\rho}_s(\mathbf{r}) \ln \left[\frac{\sin\left(p_s |\nabla_\mathbf{r}\psi(\mathbf{r})|\right)}
{p_s |\nabla_\mathbf{r}\psi(\mathbf{r})|}\right]\right ] \prod_{\mathbf{r}}\mathbf{\delta}
\left[\frac{\hat{\rho}_{p}(\mathbf{r})}{\rho_{po}} + \frac{\hat{\rho}_{s}(\mathbf{r})}{\rho_{so}} - 1\right]\label{eq:parti_elec_added}
\end{eqnarray}

where
\begin{eqnarray}
       \frac{F_a}{k_B T} & = & n_{B^+}^a\ln \mbox{K}_{B^+} - (nN-n_{H^+}^a)\ln \mbox{K}_{H^+}
- \ln \left[\frac{nN!}{n_{H^+}^a! n_{B^+}^a!(nN- n_{H^+}^a - n_{B^+}^a)!}\right] \nonumber \\
&& + \ln \left[n_{H^+}^f! n_{B^+}^f! n_s!\right] - (nN + n_s) \ln 4\pi
\end{eqnarray}
 and we have absorbed numerical prefactors of $4\pi$ coming from orientational degrees of 
freedom of the dipoles in the definitions of $\mbox{K}_{H^+}$ and $\mbox{K}_{B^+}$. Also, 
\begin{eqnarray}
       \exp\left[- i \psi_p(\mathbf{r})\right] & = & (1 - \beta_{H^+} - \beta_{B^+}) \exp\left[-i Z_p \psi(\mathbf{r})\right] 
+ \sum_{\gamma = H^+,B^+} \beta_{\gamma} \left[\frac{\sin\left(p_\gamma |\nabla_\mathbf{r}\psi(\mathbf{r})|\right)}
{p_\gamma |\nabla_\mathbf{r}\psi(\mathbf{r})|} \right]
\end{eqnarray}

Now, we introduce collective variables corresponding to $\hat{\rho}_p$ and $\hat{\rho}_s$ by using the identity 
\begin{eqnarray}
Z\left\{\hat{\rho}_p,\hat{\rho}_s\right\} &=& \int D\left[\rho_p\right] \int D\left[\rho_s\right] 
Z\left\{\rho_p, \rho_s\right\}\prod_{\mathbf{r}}\mathbf{\delta}\left[\rho_p(\mathbf{r}) - \hat{\rho}_p(\mathbf{r})\right]\left[\rho_s(\mathbf{r}) - \hat{\rho}_s(\mathbf{r})\right]  \label{eq:parti_2}
\end{eqnarray}
Also, the local 
incompressibility constraint allows us to
rewrite $H_w$ given by Eq.~\ref{eq:dispersion} in terms of collective variables as
\begin{eqnarray}
H_w &=& \frac{1}{2}\left[w_{pp} \rho_{po}nN + w_{ss}\rho_{so}n_s\right] + \chi_{ps}\int d\mathbf{r} \rho_p(\mathbf{r}) \rho_s(\mathbf{r}),\label{eq:dispersion_chi}
\end{eqnarray}
where we have used $\int d\mathbf{r} \hat{\rho}_p(\mathbf{r}) = nN,                                           
\int d\mathbf{r} \hat{\rho}_s(\mathbf{r})= n_s$. Also, we have defined a parameter $\chi_{ps}$ by
\begin{eqnarray}
\chi_{ps} &=& w_{ps} - \frac{w_{pp}\rho_{po}}{2\rho_{so}} -  \frac{w_{ss}\rho_{so}}{2\rho_{po}} \label{eq:chi_parameter}
\end{eqnarray}
which has the dimensions of volume. 

Now, writing the local constraints (represented by delta functions) in terms of functional integrals by the 
identities
\begin{eqnarray}
\prod_{\mathbf{r}}\mathbf{\delta}\left[\rho_j(\mathbf{r}) - \hat{\rho}_j(\mathbf{r})\right] 
 &=& \int D\left[w_j\right] \exp\left[-i \int d\mathbf{r} w_j(\mathbf{r})\left\{\rho_j(\mathbf{r})- \hat{\rho}_j(\mathbf{r})\right\}\right] \quad \mbox{for}\, j = p,s\label{eq:order_field} \nonumber \\
&&
\end{eqnarray}
and computing trivial functional integrals over $\rho_s(\mathbf{r})$, we arrive at Eq.~\ref{eq:parti_physical}. 
Furthermore, note that in the absence of dipolar species (i.e., $p_{j} = 0$ for $j = H^+,B^+,s$ 
in the above treatment), 
we recover the self-consistent field theory for ``annealed'' charge distributions documented in 
literature\cite{shi_scft,wang_scft}. 
 
\section*{REFERENCES}
\setcounter {equation} {0}
\pagestyle{empty} \label{REFERENCES}


\begin{thebibliography}{99}

\bibitem{alexander_brush}
S. Alexander, Journal De Physique 38 (8), 983 (1977).

\bibitem{degennes_brush}
P. G. Degennes, Macromolecules 13 (5), 1069 (1980).

\bibitem{milner_review}
S. T. Milner, Science 251 (4996), 905 (1991).

\bibitem{brush_review}
J. Ruhe, M. Ballauff, M. Biesalski, P. Dziezok, F. Grohn, D. Johannsmann, N. Houbenov, N. Hugenberg, R. Konradi, S. Minko, M. Motornov, R. R. Netz, M. Schmidt, C. Seidel, M. Stamm, T. Stephan, D. Usov, and H. N. Zhang, in Polyelectrolytes with Defined Molecular Architecture I (2004), Vol. 165, pp. 79.

\bibitem{ballauff_review}
M. Ballauff and O. Borisov, Current Opinion in Colloid \& Interface Science 11 (6), 316 (2006).

\bibitem{ballauff_review_spherical}
M. Ballauff, Progress in Polymer Science 32 (10), 1135 (2007).






\bibitem{misra_varanasi_1}
S. Misra, S. Varanasi, and P. P. Varanasi, Macromolecules 22 (11), 4173 (1989).

\bibitem{misra_varanasi_finite_extension}
S. Misra and S. Varanasi, Journal of Chemical Physics 95 (3), 2183 (1991).

\bibitem{ross_pincus}
R. S. Ross and P. Pincus, Macromolecules 25 (8), 2177 (1992).

\bibitem{zhulina_ionizable}
 E. B. Zhulina, T. M. Birshtein, and O. V. Borisov, Macromolecules 28 (5), 1491 (1995).

\bibitem{birshtein_finite_extension}
Y. V. Lyatskaya, F. A. M. Leermakers, G. J. Fleer, E. B. Zhulina, and T. M. Birshtein, Macromolecules 28 (10), 3562 (1995).

\bibitem{biesheuvel_ionizable}
P. M. Biesheuvel, Journal of Colloid and Interface Science 275 (1), 97 (2004).

\bibitem{szleifer_jps}
R. Nap, P. Gong, and I. Szleifer, Journal of Polymer Science Part B-Polymer Physics 44 (18), 2638 (2006).

\bibitem{genzer_macro}
T. Wu, P. Gong, I. Szleifer, P. Vlcek, V. Subr, and J. Genzer, Macromolecules 40 (24), 8756 (2007).

\bibitem{szleifer_macro_07}
P. Gong, T. Wu, J. Genzer, and I. Szleifer, Macromolecules 40 (24), 8765 (2007).

\bibitem{szleifer_prl}
P. Gong, J. Genzer, and I. Szleifer, Physical Review Letters 98 (1) (2007).

\bibitem{orland_scft_brushes}
H. Seki, Y. Y. Suzuki, and H. Orland, Journal of the Physical Society of Japan 76 (10) (2007).

\bibitem{kilbey07}
A.Y. Sankhe, S.M. Hussan, and S.M. Kilbey,  Journal of Polymer Science Part A - Polymer Chemistry 45(4), 566 (2007).

\bibitem{kilbey08}
S.B. Rahane, J.A. Floyd, A.T. Metters, and S.M. Kilbey, Advanced Functional Materials 18 (8), 1232 (2008).

\bibitem{witte_jpcb}
K. N. Witte, S. Kim, and Y. Y. Won, Journal of Physical Chemistry B 113 (32), 11076 (2009).

\bibitem{szleifer_pnas}
M. Tagliazucchi, M. O. de la Cruz, and I. Szleifer, Proceedings of the National Academy of Sciences of the United States of America 107 (12), 5300 (2010).

\bibitem{szleifer_langmuir}
M. J. Uline, Y. Rabin, and I. Szleifer, Langmuir 27 (8), 4679 (2011).

\bibitem{mandel_jenard1}
M. Mandel and A. Jenard, Transactions of the Faraday Society 59 (489), 2158 (1963).

\bibitem{mandel_jenard2}
M. Mandel and A. Jenard, Transactions of the Faraday Society 59 (489), 2170 (1963).

\bibitem{mandel_dielectric}
M. Mandel and T. Odijk, Annual Review of Physical Chemistry 35, 75 (1984).

\bibitem{colby_dielectric}
F. Bordi, C. Cametti and R.H. Colby, Journal of Physics - Condensed Matter 16 (49), R1423 (2004).

\bibitem{minakata}
A. Minakata, N. Imai and F. Oosawa, Biopolymers 11(2), 347 (1972).

\bibitem{debye_book}
P. Debye, {\it Polar Molecules} (The Chemical Catalog Company Inc., New York, 1929).

\bibitem{onsager_moments}
L. Onsager, Journal of Chemical Physics, 58, 1486-1493 (1936).

\bibitem{dielectric}
C.J.F. B\"{o}ttcher, {\it Theory of Electric Polarization} (Elsevier, Amsterdam, 1973).

\bibitem{booth_1}
F. Booth, Journal of Chemical Physics 19 (4), 391 (1951).

\bibitem{booth_2}
F. Booth, Journal of Chemical Physics 23 (3), 453 (1955).

\bibitem{fulton_approaches}
R. L. Fulton, Journal of Chemical Physics 130 (20), 204503 (2009).

\bibitem{sandberg}
L. Sandberg and O. Edholm, Journal of Chemical Physics 116 (7), 2936 (2002).

\bibitem{freed_dielec1}
A.K. Jha and K.F. Freed, Journal of Chemical Physics 128 (3), 034501 (2008).

\bibitem{freed_dielec2}
H. Gong and K.F. Freed, Physical Review Letters 102 (5), 057603 (2009).






\bibitem{manning}
G.S. Manning, Journal of Chemical Physics 51 (3), 924 (1969).

\bibitem{prabhu05}
V.M. Prabhu, Current Opinion in Colloid Interface Science 10 (1-2), 2 (2005).

\bibitem{levin_review}
Y. Levin, Reports on Progress in Physics 65 (11), 1577 (2002).

\bibitem{granick}
S. Wang, S. Granick and J. Zhao, Journal of Chemical Physics 129 (24), 241102 (2008).

\bibitem{kumar_condensation}
M. Muthukumar, Journal of Chemical Physics 120 (19), 9343; R. Kumar, A. Kundagrami and M. Muthukumar, Macromolecules 42 (4), 1370 (2009). 

\bibitem{edwardsbook}
M. Doi and S.F. Edwards, \textit{The Theory of Polymer Dynamics} (Clarendon Press, Oxford, 1986).

\bibitem{fredbook}
G.H. Fredrickson, {\it The Equilibrium Theory of Inhomogeneous Polymers} (Oxford University, New York, 2006).

\bibitem{intermolecular_forces}
J.N. Israelachvili, \textit{Intermolecular and Surface Forces} (Academic Press: San Diego, CA, 1987).

\bibitem{mcquarie}
D.A. McQuarie, \textit{Statistical Mechanics}, (University Science Books, Sausalito, CA, 2000).




\bibitem{muller_algo}
M. Muller, Physical Review E 65 (3), 030802 (2002).

\bibitem{numerics_imex}
U. M. Ascher, S. J. Ruuth, and B. T. R. Wetton, Siam Journal on Numerical Analysis 32 (3), 797 (1995).

\bibitem{numerics_badalassi}
V. E. Badalassi, H. D. Ceniceros, and S. Banerjee, Journal of Computational Physics 190 (2), 371 (2003).

\bibitem{handbook}
J. Wen in \textit{Polymer Data Handbook} (Oxford University Press Inc., 1999).

\bibitem{stability_constants1}
L.G. Sillen and A.E. Martell, \textit{Stability Constants of Metal-Ion Complexes: Suppl. 1} (Alden: Oxford, 1964). 

\bibitem{stability_constants2}
H.P. Gregor and M. Frederick, Journal of Polymer Science 23 (103), 451 (1957). 

\bibitem{arai_eisenberg}
K. Arai and A. Eisenberg, Journal of Macromolecular Science, Part B: Physics 17(4), 803 (1980).

\bibitem{kuhn_pmaa}
C. Ortiz and G. Hadziioannou, Macromolecules 32 (3), 780 (1999). 

\bibitem{ninham_work1}
B.W. Ninham and V. Yaminsky, Langmuir 13 (7), 2097 (1997).

\bibitem{ninham_work2}
M. Bostrom and B. W. Ninham, J. Phys. Chem. B 108 (33), 12593 (2004).

\bibitem{onuki_kitamura}
A. Onuki and H. Kitamura, Journal of Chemical Physics 121 (7), 3143 (2004).

\bibitem{wang_solvation}
Z. G. Wang, Journal of Physical Chemistry B 112 (50), 16205 (2008).

\bibitem{sadakane_1}
K. Sadakane, H. Seto and M. Nagao, Chemical Physics Letters 426 (1-3), 61 (2006).

\bibitem{sadakane_2}
K. Sadakane, H. Seto, H. Endo and M. Kojima, Journal of Applied Crystallography 40 (1), S527 (2007).

\bibitem{sadakane_3}
K. Sadakane, H. Seto, H. Endo and M. Shibayama, Journal of the Physical Society of Japan 76 (11), 113602 (2007).

\bibitem{sadakane_4}
K. Sadakane, A. Onuki, K. Nishida, S. Koizumi and H. Seto, Physical Review Letters 103 (16), 167803 (2009).

\bibitem{nabutovskii_1}
V. M. Nabutovskii, N. A. Nemov and Y. G. Peisakhovich, Physics Letters A 79 (1), 98 (1980).

\bibitem{nabutovskii_2}
V. M. Nabutovskii, N. A. Nemov and Y. G. Peisakhovich, Molecular Physics 54 (4), 979 (1985).

\bibitem{wang_solvation2}
Z.G. Wang, Physical Review E 81 (2), 021501 (2010). 

\bibitem{hansen}
D. Coslovich, J. Hansen, and G. Kahl, Journal of Chemical Physics 134 (24), 244514 (2011).

\bibitem{rob_interfacial}
R. Riggleman, R. Kumar, and G.H. Fredrickson, Journal of Chemical Physics 136 (2), 024903 (2012).

\bibitem{marcus}
R.A. Marcus, Journal of Chemical Physics 24 (5), 979 (1956).

\bibitem{felderhof}
B.U. Felderhof, Journal of Chemical Physics 67 (2), 493 (1977).

\bibitem{coalson}
R.D. Coalson and A. Duncan, Journal of Physical Chemistry 100 (7), 2612 (1996).

\bibitem{orland_dipolar}
A. Abrashkin, D. Andelman, and H. Orland, Physical Review Letters 99 (7), 077801 (2007).

\bibitem{wang08}
Z.G. Wang, Journal of Theoretical and Computational Chemistry 7 (3), 397 (2008).

\bibitem{harris_rice}
F. E. Harris, S. A. Rice, Journal of Physical Chemistry 58 (9), 725 (1954).

\bibitem{borukhov}
I. Borukhov, D. Andelman, and H. Orland, European Physical Journal B 5 (4), 869 (1998).

\bibitem{shi_scft}
A. Shi and J. Noolandi, Macromolecular Theory and Simulations 8 (3), 214 (1999).

\bibitem{wang_scft}
Q. Wang, T. Taniguchi and G.H. Fredrickson, Journal of Physical Chemistry B 108 (19), 6733 (2004).






\end{thebibliography}
\end{document}